\documentclass{article}[12pt]
\usepackage[utf8]{inputenc} 
\usepackage[T1]{fontenc}    
\usepackage{hyperref}       
\usepackage{url}            
\usepackage{booktabs}       
\usepackage{amsfonts}       
\usepackage{nicefrac}       
\usepackage{microtype}      
\usepackage{lipsum}
\usepackage{yhmath}
\usepackage{amsmath}
\usepackage{amssymb}
\usepackage{graphicx}
\usepackage{float}
\usepackage{caption}
\usepackage[english]{babel}
\usepackage{booktabs,caption}
\usepackage[flushleft]{threeparttable}
\usepackage[utf8]{inputenc}
\usepackage[noend]{algpseudocode}
\usepackage[utf8]{inputenc}
\usepackage[english]{babel}
\usepackage{amsmath, nccmath}
\usepackage{multirow}
\usepackage{geometry}
\usepackage{xcolor}
\usepackage{tabularx}
\newtheorem{theorem}{Theorem}

\DeclareMathOperator*{\argmin}{arg\,min}
\usepackage[linesnumbered,ruled,vlined]{algorithm2e}

\SetCommentSty{mycommfont}
\usepackage[
    backend=biber,
    style=nejm,
  ]{biblatex}

\SetKwInput{KwInput}{Input}                
\SetKwInput{KwOutput}{Output} 

\usepackage{graphicx}

\newcommand{\bigCI}{\mathrel{\text{\scalebox{1.07}{$\perp\mkern-10mu\perp$}}}}

\providecommand{\keywords}[1]{\textbf{\textit{Keywords---}} #1}

\title{Robust Estimation of Heterogeneous Treatment Effects using Electronic Health Record Data}

\author{
  {Ruohong Li \footnote{Department of Biostatistics and Health Data Science, Indiana University School of Medicine and 
  Fairbanks School of Public Health. Email: rhli@iu.edu}} 
   \and
 {Honglang Wang \footnote{Department of Mathematical Sciences,
  Indiana University-Purdue University Indianapolis. Email: hlwang@iupui.edu}} 
    \and
 {Wanzhu Tu \footnote{Department of Biostatistics and Health Data Science,
  Indiana University School of Medicine and
  Fairbanks School of Public Health.
  Email: wtu1@iu.edu}} 
}
\date{}
\begin{document}
\maketitle

\begin{abstract}

Estimation of heterogeneous treatment effects is an essential component of precision medicine. Model and algorithm-based methods have been developed within the causal inference framework to achieve valid estimation and inference. Existing methods such as the  A-learner, R-learner, modified covariates method (with and without efficiency augmentation), inverse propensity score weighting, and augmented inverse propensity score weighting have been proposed mostly under the square error loss function. The performance of these methods in the presence of data irregularity and high dimensionality, such as that encountered in electronic health record (EHR) data analysis, has been less studied. In this research, we describe a general formulation that unifies many of the existing learners through a common score function. The new formulation allows the incorporation of least absolute deviation (LAD) regression and dimension reduction techniques to counter the challenges in EHR data analysis. We show that under a set of mild regularity conditions, the resultant estimator has an asymptotic normal distribution. Within this framework, we proposed two specific estimators for EHR analysis based on weighted LAD with penalties for sparsity and smoothness simultaneously. Our simulation studies show that the proposed methods are more robust to outliers under various circumstances. We use these methods to assess the blood pressure-lowering effects of two commonly used antihypertensive therapies.

\end{abstract}


\section{Introduction}
The ultimate goal of precision medicine is to optimize therapeutic outcomes by tailoring medical treatment and care provision according to individual patient characteristics. In practice, such tailoring must be guided by causal treatment effects expressed as functions of the observed patient characteristics $\mathbf{x}$ \supercite{gabriel2012getting}, which account for patient heterogeneity in a given clinical population. But in reality, the true treatment effect function $\tau_0(\mathbf{x})$ is almost never known and cannot be easily ascertained from clinical trials. 

There is a sizable literature on the estimation of treatment effects in the form of $\tau_0(\mathbf{x})$. With covariates averaged out, $\tau_0(\mathbf{x})$ is reduced to the average treatment effect (ATE) $\tau_0=\int \tau_0(\mathbf{x})f(\mathbf{x})d\mathbf{x}$, which can be estimated from clinical trials as well as observational studies  \supercite{imai2008misunderstandings}. While randomized experiments provide by far the most straightforward estimation of $\tau_0$, valid estimates can also be ascertained from observational data, by using the Neyman-Rubin causal model under appropriate assumptions \supercite{sekhon2008neyman}. {Estimating treatment effect in the presence of heterogeneity, however, is  a much involved task. Popular approaches include the \emph{advantage} or A-learning methods that directly model the contrasts among treatments \supercite{murphy2003optimal,robins2004optimal}, and the \emph{quality} or Q-learners that regress the outcomes on patient characteristics  \supercite{watkins1989learning,watkins1992q}. Under the general umbrella of A-learners, Tian, Chen, and colleagues described a covariate-modification method  \supercite{tian2014simple,chen2017general}. More recently, Nie and Wager proposed a two-step learning algorithm that possesses a quasi-oracle property for estimating  $\tau_0(\mathbf{x})$ \supercite{nie2017quasi}. Xiao and colleagues further improved the algorithm for enhanced robustness \supercite{xiao2019robust}.}

The performance of the above causal estimators is often influenced by the features of the observed data. An attractive and readily available data source for causal inference is electronic health records (EHR), digitalized medical records collected and maintained by health care organizations \supercite{gunter2005emergence}. While statisticians have long recognized the values of EHR data in causal analysis \supercite{stuart2013estimating}, they are also keenly aware of the challenges presented by such data, including data outliers and  high dimensionality. The former could result in biased estimation and questionable inference, whereas the latter leads to a ``curse of dimensionality’’ \supercite{donoho2000high}.

In this research, we address the above issues in a broader context of heterogeneous treatment effect estimation. { Specifically, we put forward a general estimation framework based on  weighted score equations. The new formulation unifies many of the existing learners, while retaining the flexibility to accommodate different loss functions, permitting for example  robust least absolute deviation (LAD) regression. The estimating formula enhances modified-covariate method's capacity against outliers \supercite{chen2017general} and extends the robust R-learner's ability to handle higher dimensionality \supercite{xiao2019robust}, giving each an improvement. The approach's direct targeting of $\tau_0(\mathbf{x})$ relates it nicely to the concept of the A-learning methods.} We performed extensive simulation studies to investigate the new methods' operational performance, in comparison with the existing ones. We also described a real data application to illustrate the use of the proposed methods.

\section{Proposed Methods}
\subsection{Models and Assumptions}
We consider the estimation of $\tau_0(\mathbf{x})$, the conditional average treatment effect (CATE), within the Neyman-Rubin potential outcome framework \supercite{rubin1974estimating}.  The binary treatment indicator $T$ takes values $1$ or $-1$, i.e., $T\in\{\pm 1\}$. We let $Y^{(1)}$ and $Y^{(-1)}$ be the  \emph{potential} outcomes under $T=1$ and $T=-1$, respectively. We assume that data $\{(Y_i,T_i,\mathbf{X}_i)\}_{i=1}^n$ are independent and identically distributed (i.i.d.), where the pre-treatment covariates $\mathbf{X}_i$ could be high dimensional as in EHR analyses.  We require the stable unit treatment value assumption (SUTVA) \supercite{cox1958interpretation,rosenthal1968pygmalion} and write the \emph{observed} outcome as $Y=I(T=1)Y^{(1)}+I(T=-1)Y^{(-1)}$, where $I(\cdot)$ is an indicator function.  

Within this framework, we focus on 
\begin{align}\label{cate}
\tau_0(\mathbf{x})&=E[Y^{(1)}-Y^{(-1)}|\mathbf{X}=\mathbf{x}]=E[Y|\mathbf{X}=\mathbf{x},T=1]-E[Y|\mathbf{X}=\mathbf{x},T=-1]\\
&=\mu_1(\mathbf{x})-\mu_{-1}(\mathbf{x}),\nonumber
\end{align}
where the last part comes from the ignorability assumption defined below. This makes CATE estimation possible when $\mathbf{X}$ contains all confounders. When $T\in\{\pm 1\}$, we can always express the conditional mean outcome as
$$E(Y|\mathbf{X},T)=b_0(\mathbf{X})+\frac{T}{2}\tau_0(\mathbf{X}),$$
where $b_0(\mathbf{x})=\frac{1}{2}(E[Y^{(1)}|\mathbf{X}=\mathbf{x}]+E[Y^{(-1)}|\mathbf{X}=\mathbf{x}])$. 
This leads to a general interaction model
\begin{equation}\label{model}
    Y_i=b_0({\mathbf{X}_i})+\frac{T_i}{2}\tau_0({\mathbf{X}_i})+\varepsilon_i,
\end{equation}
where $\varepsilon_i$ is subject to Assumption~3 below, along with the other assumptions stipulated by Rubin and Rosenbaum \supercite{rubin1974estimating,rosenbaum1983central}. 

{In the existing literature, $\tau_0(\mathbf{x})$ is often depicted by a simple parametric model  \supercite{tian2014simple,xiao2019robust}. With $\mu(\mathbf{x})=E[Y|\mathbf{X}=\mathbf{x}]=b_0(\mathbf{x})+\frac{p(\mathbf{x})-1}{2}\tau_0(\mathbf{x})$, one has $Y_i-\mu(\mathbf{X}_i)=\frac{T_i-2p(\mathbf{X}_i)+1}{2}\tau_0(\mathbf{X}_i)+\varepsilon_i$, which is exactly the Robinson decomposition used by the R-learner \supercite{nie2017quasi}.}

\emph{Assumption 1 (Ignorability)}  Treatment assignment $T_i$ is independent of the potential outcomes $(Y_i^{(1)},Y_i^{(-1)})$ given the covariates $\mathbf{X}_i$, i.e., $\{Y_i^{(1)},Y_i^{(-1)} \bigCI T_i|\mathbf{X}_i\}$.

\emph{Assumption 2 (Positivity)} The propensity score $p(\mathbf{x}):=P(T=1|\mathbf{X}=\mathbf{x})\in (0,1)$. 

\emph{Assumption 3 (Conditional Independence Error)} The error is independent of the treatment assignment conditional on covariates, i.e. $\{\varepsilon_i \bigCI T_i|\mathbf{X}_i\}$. We further assume that the conditional expectation of error exists.

\subsection{A Unified Formulation for Heterogeneous Treatment Effect Estimation}

{There are two general strategies for estimating $\tau_0(\mathbf{x})$ in (\ref{model}). The first is to depict the conditional mean function $\mu_t(\mathbf{x})=E[Y|\mathbf{X}=\mathbf{x},T=t]$ with a regression model and then obtain the treatment effect estimator $\hat\tau(\mathbf{x})=\hat{\mu}_1(\mathbf{x})-\hat{\mu}_{-1}(\mathbf{x})$. For example, from the objective function
$\sum_{i=1}^n \rho\left(Y_i-b(\mathbf{X}_i;\boldsymbol\gamma)-\frac{T_i}{2}\tau(\mathbf{X}_i;\boldsymbol\beta)\right),$ one can estimate $\boldsymbol\beta$ and $\boldsymbol\gamma$ simultaneously, and then achieves a CATE estimate  $\hat{\tau}(\mathbf{x})=\tau(\mathbf{x};\hat{\boldsymbol\beta})$ \supercite{chakraborty2013statistical}. Such an approach is often referred to as the Q-learning, because its objective function plays a role similar to that of the Q or reward function in reinforcement learning \supercite{chakraborty2013statistical}. The frequently used Two- or Single-learners (T or S-learners for short) are variants of this approach \supercite{kunzel2019metalearners}. }

{An alternative strategy, one that we follow in the current research, is to directly target $\tau_0(\mathbf{x})$ in a predefined objective function. This approach is often referred to as the A-learning \supercite{schulte2014q}. A-learning first emerged in the context of dynamic treatment regime \supercite{murphy2003optimal,robins2004optimal}, and was later generalized to one-stage case for treatment effect estimation  \supercite{tian2014simple,chen2017general}. In this paper, we show that there exists a unified formulation for the objective function, written in the form of score equations, that covers many of the existing learners. }

{Before introducing the general formulation, we first review the existing methods to highlight their connections.}

1. \emph{The modified outcome methods.} Certain transformations of $Y$ could be used to facilitate the estimation of $\tau_0(\mathbf{x})$. Estimation methods relying on such transformations are collectively known as the {modified outcome methods}. This class of methods includes the \emph{inverse propensity score weighting} (IPW) \supercite{horvitz1952generalization,hirano2003efficient} and the \emph{augmented IPW} (AIPW) methods \supercite{robins1995semiparametric}. A common feature of this class of methods is to express the true treatment effect $\tau_0(\mathbf{x})$ as a conditional expectation of the \emph{transformed} outcome variables. For IPW and AIPW, the transformations are
$$\begin{aligned}
Y^{IPW}&=\frac{T-2p(\mathbf{X})+1}{2p(\mathbf{X})(1-p(\mathbf{X}))}\times Y;\\
Y^{AIPW}&=\frac{T-2p(\mathbf{X})+1}{2p(\mathbf{X})(1-p(\mathbf{X}))}\times[Y-(\mu_{-1}(\mathbf{X})p(\mathbf{X})+\mu_{1}(\mathbf{X})(1-p(\mathbf{X})))].
\end{aligned}$$
Writing the modified outcome as $Y^*$ one has $E(Y^*|\mathbf{X})=\tau_0(\mathbf{X})$. An estimate can therefore be obtained by minimizing the square error loss, i.e.,  $\min_{\tau(\cdot)}\sum_{i=1}^n (Y_i^*-\tau(\mathbf{X}_i))^2.$

2. \emph{The modified covariates methods.} An alternative set of methods, collectively known as the modified covariates methods, have been derived from the model (\ref{model}). The central idea of this approach is to estimate $\tau_0(\mathbf{X})$ by re-weighting the loss function instead of the response variable \supercite{tian2014simple,chen2017general}
$$L(\tau(\cdot))=\sum_{i=1}^n \left(D_i\frac{1}{p(\mathbf{X}_i)}+(1-D_i)\frac{1}{1-p(\mathbf{X}_i)}\right)\left(Y_i-\frac{T_i}{2}\tau(\mathbf{X}_i)\right)^2,$$
where $D_i=(T_i+1)/{2} \in \{0,1\}$. With appropriate weighting, the minimizer of the population version of the objective function equals to $\tau_0(\mathbf{x})$ as elaborated in Remark 1 below. Further, as shown in Appendix A.1, $Y_i$ can be replaced by $Y_i-g(\mathbf{X}_i)$, where $g(\mathbf{X}_i)$ is an arbitrary function of $\mathbf{X}_i$. When $p(\mathbf{X}_i)=\frac{1}{2}$,  the variance of the estimator is  minimized when we replace $Y_i$ with $Y_i-\mu(\mathbf{X}_i)$. This is known as the \emph{modified covariates method with efficiency augmentation (MCM-EA)} \supercite{tian2014simple}.

3. \emph{The R-learning method}.  Nie and Wager recently proposed a method that they referred to as the R-learner (RL) \supercite{nie2017quasi}, named after Robinson's decomposition, a technique for estimating the parametric components in partially linear models \supercite{robinson1988root}. The efficient A-learning introduced later in this section shared the same estimating equation of the R-learner, but the two were derived from different perspectives \supercite{robins2004optimal,lu2013variable}. Subtracting the marginal mean $E[Y_i|\mathbf{X}_i]$ from the outcome, Nie and Wager worked with the following equation 
$$Y_i-E[Y_i|\mathbf{X}_i]=\left(\frac{T_i}{2}-p(\mathbf{X}_i)+\frac{1}{2}\right)\tau_0(\mathbf{X}_i)+\varepsilon_i,$$
where $E[\varepsilon_i|\mathbf{X}_i,T_i]=0$. The treatment effect $\tau_0(\mathbf{x})$ can therefore be estimated by minimizing the following objective function, 
$$L(\tau(\cdot))=\sum_{i=1}^n \left(Y_i-\mu(\mathbf{X}_i)-\frac{T_i-2p(\mathbf{X}_i)+1}{2}\tau(\mathbf{X}_i)\right)^2,$$ where $\mu(\mathbf{X}_i)$ and $p(\mathbf{X}_i)$ are nuisance quantities estimated in advance. 

{Examining the relations between MCM-EA and R-learner, we note} that in MCM-EA, since $E[Y_i-\mu(\mathbf{X}_i)-\frac{T_i}{2}\tau_0(\mathbf{X}_i)|\mathbf{X}_i]\neq 0$, one uses IPW as an adjustment so that $E[\frac{T_i}{2T_i p(\mathbf{X}_i)+(1-T_i)}\left(Y_i-\frac{T_i}{2}\tau_0(\mathbf{X}_i)\right)|\mathbf{X}_i]=0$. In the R-learning, one has $E[Y_i-\mu(\mathbf{X}_i)-(\frac{T_i}{2}-p(\mathbf{X}_i)+\frac{1}{2})\tau_0(\mathbf{X}_i)|\mathbf{X}_i]=0$, so propensity score adjustment becomes unnecessary. This shows the difference and the connection between the R-learning and MCM-EA. 

{4. \emph{The A-learning (AL) methods}. {By directly targeting at the contrast function (treatment effect function), Robins \supercite{robins2004optimal} derived the following equation for CATE estimation,} 
$$E\left[ \left(Y_i-\theta(\mathbf{X}_i)-\frac{T_i+1}{2}\tau_0(\mathbf{X}_i)\right) \left(T_i-2p(\mathbf{X}_i)+1\right)\Bigg|\mathbf{X}_i\right]=0,$$
where $\theta(\cdot)$ is an arbitrary function, or a more efficient version
\begin{equation}\label{a-learning2}
E\left[ \left(Y_i-\mu(\mathbf{X}_i)-\frac{T_i-2p(\mathbf{X}_i)+1}{2}\tau_0(\mathbf{X}_i)\right) \left(T_i-2p(\mathbf{X}_i)+1\right)\Bigg|\mathbf{X}_i\right]=0,
\end{equation}
where the first term $Y_i-\mu(\mathbf{X}_i)-\frac{T_i-2p(\mathbf{X}_i)+1}{2}\tau_0(\mathbf{X}_i)$  has mean $0$ conditional on $\mathbf{X}_i$. This  corresponds exactly to Robinson's decomposition $Y_i-\mu(\mathbf{X}_i)=\frac{T_i-2p(\mathbf{X}_i)+1}{2}\tau_0(\mathbf{X}_i)+\epsilon_i$ when $E(\epsilon_i|\mathbf{X}_i)=0$.} Note that $Y_i-\mu(\mathbf{X}_i)-\frac{T_i-2p(\mathbf{X}_i)+1}{2}\tau_0(\mathbf{X}_i)=Y_i-\mu_{-1}(\mathbf{X}_i)-\frac{T_i+1}{2}\tau_0(\mathbf{X}_i)$ since $\mu(\mathbf{x})=\mu_{-1}(\mathbf{x})+p(\mathbf{x})\tau(\mathbf{x})$. The $\mu_{-1}(\mathbf{x})$ version was used by several authors \supercite{schulte2014q,tsiatis2019dynamic}.

 {This shows that Nie's R-learner shares the same conceptual essence with Robin's efficient A-learner, although the two were derived from different perspectives. }

{In summary,  the methods reviewed above, including IPW, AIPW, MCM, MCM-EA, and RL could all be viewed as variants of AL,  since they all target $\tau_0(\cdot)$ directly, {with the pre-estimated plug-in nuisance quantities}. We now show that these methods can be formulated under a unified presentation of the objective functions, at the level of score equations. }  

{Noting that the above learners are all based on solutions to some score equations corresponding to the objective functions under the square error loss, we specify the score equations for these methods:}
\begin{itemize}
\item Modified covariates: $S_{MCM}=\frac{T}{2T p(\mathbf{X})+(1-T)}\left(Y-\frac{T}{2}\tau_0(\mathbf{X})\right)$;
\item Modified covariates with efficiency augmentation: $S_{MCM-EA}=\frac{T}{2T p(\mathbf{X})+(1-T)}\left(Y-\mu(\mathbf{X})-\frac{T}{2}\tau_0(\mathbf{X})\right)$;
\item R learning (efficient A learning): $S_{RL}=\frac{T-2p(\mathbf{X})+1}{2}\left(Y-\mu(\mathbf{X})-\frac{T-2p(\mathbf{X})+1}{2}\tau_0(\mathbf{X})\right)$;
\item Inverse probability weighting: $S_{IPW}= \frac{T-2p(\mathbf{X})+1}{2p(\mathbf{X})(1-p(\mathbf{X}))} \left(Y-\frac{2p(\mathbf{X})(1-p(\mathbf{X}))}{T-2p(\mathbf{X})+1}\tau_0(\mathbf{X})\right)$;
\item AIPW: $S_{AIPW}=\frac{T-2p(\mathbf{X})+1}{2p(\mathbf{X})(1-p(\mathbf{X}))} \left(Y-((1-p(\mathbf{X}))\mu_1(\mathbf{X})+p(\mathbf{X})\mu_{-1}(\mathbf{X}))-\frac{2p(\mathbf{X})(1-p(\mathbf{X}))}{T-2p(\mathbf{X})+1}\tau_0(\mathbf{X})\right)$.
\end{itemize}

We note that all score equations listed above can be expressed in one general formulation
\begin{equation}\label{score}
S=w(\mathbf{X},T)c(\mathbf{X},T)[Y-g(\mathbf{X})-c(\mathbf{X},T)\tau_0(\mathbf{X})],
\end{equation}
where the two weight functions $w(\mathbf{x},t)$ and $c(\mathbf{x},t)$ are subject to the following constraints \textcolor{blue}{for all $x$ and $t$}:
\begin{itemize}
\item[C1.] $p(\mathbf{x})w(\mathbf{x},1)c(\mathbf{x},1)+(1-p(\mathbf{x}))w(\mathbf{x},-1)c(\mathbf{x},-1)=0$;
\item[C2.] $c(\mathbf{x},1)-c(\mathbf{x},-1)=1$;
\item[C3.] $w(\mathbf{x},t)c(\mathbf{x},t) \ne 0$.
\end{itemize}
One can show that the existing estimation methods, including MCM, MCM-EA, RL, IPW, and AIPW, are all covered by this general formulation. In Appendix A.1, we show that for each of the above methods, {the corresponding functions $c$ and $w$ meet the three conditions.}

A few additional remarks are in order for this general expression: 

\noindent\emph{Remark 1}.  Conditions C1-C3 are put in place to assure $E(S|\mathbf{X})=0$. It can be shown that under the square error loss function, the estimates derived from (\ref{score}) are indeed minimizers of the target function, i.e.,
$\tau_0(\mathbf{x})=argmin_{\tau(\mathbf{x})}E[w(\mathbf{X}_i,T_i)(y-g(\mathbf{X}_i)-c(\mathbf{X}_i,T_i)\tau(\mathbf{x}))^2|\mathbf{X}_i=\mathbf{x}].$ For detailed proof, see Property 1 in Appendix A.2. A similar results can be obtained under the absolute error loss function; see Property 3 in the same section of the appendix.

\noindent\emph{Remark 2.} {For given $w(\mathbf{x},t)$ and $c(\mathbf{x},t)$, one might be able to choose an appropriate $g(\mathbf{x})$ to achieve robustness to model mis-specification. For example, the $g(\mathbf{x})=(1-p(\mathbf{x}))\mu_1(\mathbf{x})+p(\mathbf{x})\mu_{-1}(\mathbf{x})$ in the augmented inverse probability weighting (AIPW) method with equation
$$E\left[\frac{T_i-2p(\mathbf{X}_i)+1}{2p(\mathbf{X}_i)(1-p(\mathbf{X}_i))}\left(Y_i-g(\mathbf{X}_i)-\frac{2p(\mathbf{X_i})(1-p(\mathbf{X}_i))}{T_i-2p(\mathbf{X}_i)+1}\tau_0(\mathbf{X}_i)\right)\bigg|\mathbf{X}_i\right]=0$$
leads to double robustness. Specifically, AIPW is robust against mis-specification of either propensity score model or both $\mu_{-1}(\mathbf{x})$ and $\mu_1(\mathbf{x})$.}

\noindent\emph{Remark 3.}  When an additional condition $c(\mathbf{x},1)=1-p(\mathbf{x})$ holds and  $g(\mathbf{x})=\mu(\mathbf{x})$,  the score equation in (\ref{score}) leads to an estimator with the minimized variance. {For an R learner, we have $c(\mathbf{x},1)=1-p(\mathbf{x})$, and the choice of $g(\mathbf{x})=\mu(\mathbf{x})$ leads to the most efficient estimator.}  For MCM, this additional condition also holds when $p(\mathbf{x})=\frac{1}{2}$, as in the case of randomized clinical trials.

\noindent\emph{Remark 4.} {With the unified formulation for the score functions, new estimators can be derived, for example, $E[(T_i-2p(\mathbf{X}_i)+1)(Y_i-g(\mathbf{X}_i)-\frac{T_i}{2}\tau_0(\mathbf{X}_i))|\mathbf{X}_i]=0$, where $g(\mathbf{X})$ is an arbitrary augmented function of $\mathbf{X}$.}

With the score function expressed as in (\ref{score}), we propose an estimation procedure for CATE $\tau(\cdot)$,
\begin{equation}\label{general}
\min_{\tau(\cdot)}\frac{1}{n} \sum_{i=1}^n w(\mathbf{X}_i,T_i) \rho(Y_i-g(\mathbf{X}_i)-c(\mathbf{X}_i,T_i)\tau(\mathbf{X}_i))+\Lambda_n(\tau(\cdot)),   
\end{equation}
where $\rho(\cdot)$ is a user-specified loss function, and $\Lambda_n(\cdot)$ is a structural penalty function for $\tau(\cdot)$.  This general procedure covers most of the existing methods for heterogeneous treatment effect estimation through a unified formulation.

\subsection{Estimation Methods under the $L_1$ Loss}

The estimation procedure described in (\ref{general}) is general and flexible in the sense that it allows the analyst: (1) to choose different estimators through the specification of $w(\cdot)$ and $c(\cdot)$; (2) to select $g(\cdot)$ for efficiency enhancement; and (3) to specify a loss function $\rho(\cdot)$ that is most appropriate for the application. This general formulation provides a natural remedy to two practical issues in EHR data analysis: (1) lack of robustness of the $L_2$-based methods against outliers, (2) lack of accommodation of the high dimensionality of $\mathbf{X}$, and nonlinearity of  $\tau(\mathbf{X})$.

Specifically, we put forward a class of robust estimators within the confines of the general estimating function (\ref{general}). The method accommodates nonlinearity in $\tau(\cdot)$, further enhancing the modeling flexibility. Estimation is implemented under the usual causal inference Assumptions $1-3$.

Under the $L_1$-loss function, we show in Appendix A that with Conditions C1-C3, we have 
$$
argmin_{\tau(\cdot)}E\left[ w(\mathbf{X}_i,T_i) \cdot |Y_i-g(\mathbf{X}_i)-c(\mathbf{X}_i,T_i)\tau(\mathbf{X}_i)|\bigg|\mathbf{X}_i=\mathbf{x}\right]=\tau_0(\mathbf{x}). 
$$
To increase efficiency, we opt to use $g(\mathbf{X}_i)=\mu(\mathbf{X}_i)$ in proposed methods.

Herein, we consider the following penalized least absolute deviation estimator
\begin{equation}\label{loss}
\min_{\tau(\cdot)}\frac{1}{n} \sum_{i=1}^n w(\mathbf{X}_i,T_i) |Y_i-\mu(\mathbf{X}_i)-c(\mathbf{X}_i,T_i)\tau(\mathbf{X}_i)|+\Lambda_n(\tau(\cdot)), 
\end{equation}
where  $\Lambda_n$ is added to ensure sparsity at the function level. For simultaneous variable selection and smooth estimation, we adopt a similar penalty term in (\ref{loss}) as described by Meier \supercite{meier2009high}. 

We further assume an additive structure for the treatment effect function $\tau(\cdot)$:
$$\tau(\mathbf{x})=\alpha+m_1(x_{1})+m_2(x_{2})+...+m_p(x_{p}),$$ 
where $\alpha$ is the intercept, and $m_j(\cdot)$ is the $j$th additive component corresponding to $x_j$. We write
$$m_j(x_j)=\sum_{k=1}^{K_n+q} B_{jk}(x_j)\beta_{jk},$$
where $\{B_{jk}(x_j)\}_{k=1}^{K_n+q}$ are the B-spline basis functions, $K_n$ and $q$ are number of knots and degree.

Rewriting the spline bases and coefficients as vectors, we have $\tau(\mathbf{x})=\alpha+\boldsymbol\beta^TB(\mathbf{x})$, where $B(\mathbf{x})=(B^T_1(x_1),\cdots, B^T_p(x_p))^T=(B_{11}(x_1),B_{12}(x_1),\cdots,B_{1(K+q)}(x_1),\cdots,B_{p(K+q)}(x_p))^T$, $\boldsymbol{\beta}=(\boldsymbol\beta_1^T,\cdots,\boldsymbol\beta_p^T)^T=(\beta_{11},\beta_{12},\cdots,\beta_{1(K+q)},\cdots,\beta_{p(K+q)})^T$.  For simplicity, we choose a common $K_n+q$ for all spline components. Following a suggestion of $K_n\asymp \sqrt{n}+4$ by Meier  \supercite{meier2009high}, we use $K_n=\sqrt{n}/2$, which is of the same order and not too large for implementation. 

With this, we define the penalty term in (\ref{loss}) as 
\begin{equation}\label{penalty}
\Lambda_n(\tau(\cdot))=\sum_{j=1}^p P_{\lambda_1,\gamma}(J(m_j)), \text{ with} \  J(m_j)=\sqrt{||m_j||^2_n+\lambda_2I^2(m_j)},
\end{equation} 
{where $||m_j||^2_n=\frac{1}{n} \sum_{i=1}^n m_j^2(\mathbf{X}_i)=\frac{1}{n}\boldsymbol\beta_j^T \mathbf{D}_j \boldsymbol\beta_j$ is for variable selection in a group-wise manner, and $I^2(m_j)=\int (m_j''(\mathbf{x}))^2d\mathbf{x}=\boldsymbol\beta_j^T \mathbf{\Omega}_j \boldsymbol\beta_j$ is for smoothness of the nonzero components. } The integrals $\int B_{jl_1}(\mathbf{x})B_{jl_2}(\mathbf{x})d\mathbf{x}$ and $\int B_{jl_1}''(\mathbf{x})B_{jl_2}''(\mathbf{x})d\mathbf{x}$ are the $(l_1,l_2)$th entry of the $(K_n+q) \times (K_n+q)$ matrices $\mathbf{D}_j$ and $\mathbf{\Omega}_j$ respectively. And $P_{\lambda_1,\gamma}(\cdot)$ is the smoothly clipped absolute deviation (SCAD) penalty defined by its first derivative
$$P'_{\lambda_1,\gamma}(x)=\lambda_1\{I(x \le \lambda_1)+\frac{(\gamma \lambda_1-x)_+}{(\gamma-1)\lambda_1} I(x>\lambda_1)\},$$
with $\gamma>2$ and $P_{\lambda_1,\gamma}(0)=0$. We use $\gamma=3.7$ as suggested by Yuan and Lin \supercite{yuan2006model}.  

Hence, optimization of (\ref{loss}) can be expressed as a general group SCAD problem
$$
    (\hat{\alpha},\hat{\beta})=argmin_{(\alpha,\beta)} \frac{1}{n}\sum_{i=1}^n w(\mathbf{X}_i,T_i) \left|Y_i-g(\mathbf{X}_i)-c(\mathbf{X}_i,T_i)(\alpha+\beta^TB(\mathbf{X}_i))\right|+\sum_{j=1}^p  P_{\lambda_1,\gamma}(\sqrt{ \boldsymbol\beta_j^T M_j(\lambda_2) \boldsymbol\beta_j}),
$$
where $\mathbf{M}_j(\lambda_2)=\frac{1}{n}\mathbf{D}_j +\lambda_2 \mathbf{\Omega}_j$. By decomposing $\mathbf{M}_j=\mathbf{R}_j^T \mathbf{R}_j$ for some invertible matrix $\mathbf{R}_j \in \mathbb{R}^{(K_n+q) \times (K_n+q)}$, we define 
\begin{equation}\label{beta}
    \Tilde{\boldsymbol\beta}_j^T= \boldsymbol\beta_j^T\mathbf{R}_j
    \quad\text{and}\quad
    \Tilde{B_j}(X_j)=\mathbf{R}_j^{-1}B_j(X_j).
\end{equation}
With these transformations, the optimization of (\ref{loss}) becomes an ordinary least absolute deviation (LAD) regression with a group SCAD penalty 
\begin{equation}\label{optimization}
(\hat{\alpha},\hat{\Tilde{\boldsymbol\beta}})=argmin_{(\alpha,\Tilde{\boldsymbol\beta})} \frac{1}{n} \sum_{i=1}^n |Y_i^*-w_i^*(\mathbf{X}_i,T_i)(\alpha+\Tilde{\boldsymbol\beta}^T\Tilde{B}(\mathbf{X}_i))|+ \sum_{j=1}^p P_{\lambda_1,\gamma}(||\Tilde{\boldsymbol\beta}_j||),    
\end{equation}
where $||\Tilde{\boldsymbol\beta}_j||$ is the Euclidean norm, $Y_i^*=w(\mathbf{X}_i,T_i)(Y_i-g(\mathbf{X}_i))$ and $w_i^*(\mathbf{X}_i,T_i)=w_i(\mathbf{X}_i,T_i)c(\mathbf{X}_i,T_i)$.  The estimation of CATE is therefore $\hat{\tau}(\mathbf{x})=\hat{\alpha}+\hat{\Tilde{\boldsymbol\beta}}^T\Tilde{B}(\mathbf{x})$.

\subsection{A Computational Algorithm} 

To optimize (\ref{optimization}), one has to estimate $\mu(\cdot)$ and $p(\cdot)$, as they are involved in the weight functions $w(x,t)$ and $c(x,t)$. Herein, we use pre-estimated $\hat{\mu}(\cdot)$ and $\hat{p}(\cdot)$ as plug-in estimates for solving (\ref{optimization}).  Estimation accuracy of these quantities, however, can be impeded by the dimension of $\mathbf{x}_i$  and the uncertainty of the functional forms of the $\mathbf{x}_i$'s associations with $T_i$ and $Y_i$. To remedy, we use a gradient boosting machine (GBM)\supercite{mccaffrey2004propensity} to estimate these two functions , with packages \verb+gbm+ \supercite{ridgeway2004gbm} and \verb+caret+ \supercite{kuhn2012caret}. In cases of ultra-high dimensional $\mathbf{x}_i$, one could first use non-parametric independence screening (NIS) method \supercite{fan2011nonparametric} to reduce the dimensionality to a moderate one ($n-1$ or $log(n))$ as suggested by Fan and Lv \supercite{fan2008sure}), before applying our proposed method. 

With the plug-in estimates of $\mu(\cdot)$ and $p(\cdot)$, we solve the $L_1$ optimization problem in (\ref{optimization}), by using R package \verb+rqPen+ \supercite{sherwood2016rqpen}, which is designed for penalized quantile regression in general. The nonconvex group penalized optimization with quantile loss is solved by the extension of quantile iterative coordinate descent (QICD) algorithm proposed by Peng and Wang \supercite{peng2015iterative}. For comparison purposes, we also use R package \verb+oem+ \supercite{xiong2016orthogonalizing} to ascertain the $L_2$ estimators. 

The main steps of the procedure are described in Algorithm \ref{alg:euclid}.

\begin{algorithm}[H]
\caption{}\label{alg:euclid}
\KwInput{Outcome $Y$, treatment assignment $T$, and pre-treatment covariates $\mathbf{X}$}
\textbf{Data screening:} Screen covariates with NIS when in situations of ultra-high dimension.

\textbf{Nuisance quantity estimation:} {Estimate ${p}(\mathbf{x})$ by using GBM with cross-validation (CV) and estimate ${\mu}(\mathbf{x})$ by using $L_1$-based GBM with CV.}

{\textbf{Data transformation:} Construct the B-spline design matrix $B(\mathbf{X})$, calculate $w(\mathbf{X}_i,T_i)$, $c(\mathbf{X}_i,T_i)$, and  $g(\mathbf{X}_i)$ following Conditions C1-C3, and transform $B(\mathbf{X})$ to $\Tilde{B}(\mathbf{X})$ using (\ref{beta}).}

\textbf{Optimization:} Solve penalized LAD regression (\ref{optimization}) with a group SCAD penalty to achieve estimates of $\alpha$ and $\Tilde{\boldsymbol\beta}$ with regularization parameters selected by CV.

\KwOutput{Calculate $\hat{\tau}(\mathbf{x})=\hat{\alpha}+\hat{\Tilde{\boldsymbol\beta}}^T\Tilde{B}(\mathbf{x})$.}
\end{algorithm}

\section{Asymptotic Properties of $\hat{\tau}(x)$}

{For theoretical examination, we consider the simple case of a univariate covariate $X_i\in\mathbb{R}$:
\begin{equation}\label{simple}
\min_{\tau(\cdot)}\frac{1}{n} \sum_{i=1}^n w(X_i,T_i) \rho(Y_i-g(X_i)-c(X_i,T_i)\tau(X_i)),  
\end{equation}
where $\rho(\cdot)$ is a loss function that is convex and has unique minimizer at origin. This simplification will not diminish the contribution of the asymptotic analysis, which is complicated by the B-spline approximation and the various loss functions including the $L_1$, $L_2$, Huber, and Bisquare loss functions.}  

With a B-spline approximation, we write $\tau(x):=\sum_{k=1}^{K_n+q}\beta_kB_k(x)=B(x)^T\mathbf{\boldsymbol\beta}$, where $q$ is the degree of the B-splines and $K_n$ is the number of knots, which we assume depending on sample size $n$. Zhou et. al. \supercite{zhou1998local} provided the $L_{\infty}$ approximation error for B-splines. In particular, with $\tau^*(x):=B(x)^T\mathbf{\boldsymbol\beta}^*$ as the best $L_{\infty}$ approximation to the true function $\tau_0(x)$, it satisfies
\begin{equation}
 sup_{x \in (0,1)}|\tau^*(x)-\tau_0(x)-b^a(x)|=o(K_n^{-(q+1)}),   
\end{equation}
where 
$$b^a(x)=-\frac{\tau_0^{(q+1)}(x)}{K_n^{(q+1)}(q+1)!}\sum_{k=1}^{K_n}I(\kappa_{k-1}\le x < \kappa_k) Br_{q+1}\left(\frac{x-\kappa_{k-1}}{K_n^{-1}}\right)=O(K_n^{-(q+1)}),$$
with $\{\kappa_k\}_{k=0}^{K_n}$ are the knots in the B-spline approximation, $\tau_0^{(q+1)}(x)$ is the $(q+1)$th order derivative of $\tau_0(x)$, and $Br_q(x)$ is the q-th Bernoulli polynomial. 

We focus on the asymptotic theory of the $L_1$ spline estimator $\hat\tau(x)=B(x)^T\hat{\mathbf{\boldsymbol\beta}}$, where
{\begin{equation}\label{bspline-estimator}
\hat{\boldsymbol\beta}=\argmin_{\boldsymbol\beta\in\mathbb{R}^{K_n+q}} L_n(\boldsymbol\beta):=\sum_{i=1}^n w(X_i,T_i) \rho(Y_i-g(X_i)-c(X_i,T_i)B(X_i)^T\mathbf{\boldsymbol\beta}).  
\end{equation}
The error for $\hat{\tau}(x)$ can be decomposed as a summation of the estimation error and approximation error}
$$\hat{\tau}(x)-\tau_0(x)=\underbrace{\hat{\tau}(x)-\tau^*(x)}_\text{estimation error}+\underbrace{\tau^*(x)-\tau_0(x)}_\text{approximation error}=\hat{\tau}(x)-\tau^*(x)+b^a(x)+o(K_n^{-(q+1)}).$$
We  only need to study the estimation error $\hat{\tau}(x)-\tau^*(x)=B(x)^T(\hat{\boldsymbol\beta}-\boldsymbol\beta^*)$ thanks to the $L_{\infty}$ approximation result by Zhou et al. \supercite{zhou1998local}.

To show a pointwise asymptotic normality of $\sqrt{a_n}(\hat{\tau}(x)-\tau^*(x))$ with a convergence rate $a_n$ to be specified later in Appendix A, we only need to prove the convergence of $\sqrt{a_n}(\hat{\boldsymbol\beta}-\boldsymbol\beta^*)$ since $\hat{\tau}(x)-\tau^*(x)=B(x)^T(\hat{\boldsymbol\beta}-\boldsymbol\beta^*)$. {For this, denote $\boldsymbol\delta=\sqrt{\alpha_n}(\boldsymbol\beta-\boldsymbol\beta^*)$ and 
{$$
U_n(\boldsymbol\delta) =  \sum_{i=1}^n  \left[w(X_i,T_i)\left(\rho\left(U_i-\frac{1}{\sqrt{\alpha_n}} c(X_i,T_i) B(X_i)^T\boldsymbol\delta\right)-\rho\left(U_i\right)\right)\right],
$$
where $U_i=Y_i-g(X_i)-c(X_i,T_i)B(X_i)^T\boldsymbol\beta^*$. }}
Then the minimizer $\hat{\boldsymbol\delta}_n$ of $U_n(\boldsymbol\delta)$ is simply our target, i.e., $\hat{\boldsymbol\delta}_n=\sqrt{a_n} (\hat{\boldsymbol\beta}-\boldsymbol\beta^*).$

If one regards $\{U_n(\boldsymbol\delta)\}$ as a sequence of random functions and the finite-dimensional distributions of $U_n(\boldsymbol\delta)$ converge in distribution to those of some random function $U(\boldsymbol\delta)$ which has a unique minimum, then it will follow that 
$\hat\delta_n=\sqrt{a_n}(\hat{\boldsymbol\beta}-\boldsymbol\beta^*) \rightarrow_d argmin(U(\boldsymbol\delta)),$ as $n\rightarrow \infty$ per Hjort and Pollard \supercite{hjort1993asymptotics}, and Geyer \supercite{geyer1996asymptotics}.

With a given loss function $\rho(\cdot)$, we define $\Phi(s|X=x,T=t)=E[\rho(Y-g(x)-c(x,t)B(x)^T\boldsymbol\beta^*-s)|X=x,T=t]$. Let $\Phi'(s|X=x,T=t)$ and $\Phi''(s|X=x,T=t)$ be the first and second derivative of $\Phi(s|X=x,T=t)$ with for $\hat\delta_n$; respect to $s$. Several additional conditions are required for the proof of asymptotic normality:
\begin{itemize}
\item[C4.] $X$ is distributed as $Q(x)$ on a compact set in $\mathbb{R}$. Without loss of generality, we assume $X\in [0,1]$.
\item[C5.] The B-spline knots are equidistantly located as $\kappa_k=k/K_n, k=0,...,K_n$ and the number of knots satisfies {$K_n=O(n^{1/(2q+3)})$}.
{\item[C6.] The true CATE $\tau_0(x)$ is $(q+1)$th order continuously differentiable.
\item[C7.] The function $\rho(u)$ is convex, it has a unique minimizer at zero, and its first and second derivatives exist.
\item[C8.] For $x \in [0,1]$ and $t \in \{\pm 1\}$, $E[\rho'(Y-g(X)-c(X,T)\tau_0(X))^2|X=x,T=t] <\infty$.
\item[C9.] $\Phi(s|X=x,T=t)$, $\Phi'(s|X=x,T=t)$, and $\Phi''(s|X=x,T=t)$ are functions of $s$ and they are bounded and continuous in a neighborhood of zero.
\item[C10.] {As $s \rightarrow 0$,
$E[\{w(X,T)\big(\rho\left(U-s\right)-\rho(U)-\rho'(U)s\big)\}^2]=o(s^2).
$}
\item[C11.] There exists a $\gamma>0$ such that for any $x \in [0,1]$ and $t \in \{\pm 1\}$, $E[|w(X,T)c(X,T)\rho'(U)|^{2+\gamma}|X=x,T=t]<\infty$.}
\end{itemize}

\noindent\emph{Remark 5.} {The above conditions are needed for establishing an asymptotic normality of the estimator. Conditions C4-C6 are standard assumptions for B-spline regression. C5 provides the appropriate conditions of the knots. It suggests that the locations of the knots are set to some extent at regular intervals and the number of knots increases with the sample size. C4-6 are needed for controlling the spline approximation bias. C7-C8 are the general conditions for the loss function. The commonly used  $L_1$, Huber, and Bisquare loss functions for robust regression all satisfy these conditions. C7 also guarantees the uniqueness of the estimator. C9 and C10 ensure the smoothness of the loss function $\rho$, which are needed for controlling the remainder term in the Taylor expansion. C11 is needed for satisfying the Lyapunov condition of the Central Limit Theorem.}

To describe the asymptotic normality of the spline estimator $\hat{\tau}(x)$, { we introduce  two matrices: We define a square matrix $
\mathbf{G} \in \mathbb{R}^{(K_n+q)\times(K_n+q)}$  with $(i,j)$-th element $G_{ij}$  }
$$\mathbf{G}_{ij}=\int_0^1  \frac{p(x)}{1-p(x)}w^2(x,1)c^2(x,1)\rho'(U_i)^2 B_{i}(x)B_{j}(x) dQ(x),$$
{ and another square matrix $\mathbf{D}$ of the same dimension with its $(i,j)$-th element being}
\begin{equation*}
\begin{split}
\mathbf{D}_{ij}=&\int_0^1 \nu(x) B_{i}(x)B_{j}(x) dQ(x),
\end{split}
\end{equation*}
where $\nu(x)=p(x)w(x,1)c(x,1)^2\rho''(y^{(1)}-g(x)-c(x,1)B(x)^T\boldsymbol\beta^*)+(1-p(x))w(x,-1)c(x,-1)^2\rho''(y^{(-1)}-g(x)-c(x,-1)B(x)^T\boldsymbol\beta^*)$.

{\begin{theorem} Assuming C1-C11, as $n \rightarrow \infty$, we have
$\sqrt{n/K_n}(\hat{\tau}(x)-\tau_0(x)-b^a(x)) \stackrel{D}{\rightarrow} N(0,\Psi(x)),$
where $\Psi(x)=lim_{n \rightarrow \infty} \frac{1}{4K_n}B(x)^T\mathbf{D}^{-1}\mathbf{G}\mathbf{D}^{-1}B(x).$\end{theorem}}

\noindent\emph{Remark 6.} {With the order of $K_n$ larger than $O(n^{\frac{1}{2q+3}})$, the B-spline approximation error $b^a(x)$ can be ignored relative to the order of its variance. }

{For the rest of the paper, we focus on the LAD loss where Conditions C7-C10 are naturally satisfied, and C11 can be simplified as the following:}
{
\begin{itemize}
\item[C12.] There exists a constant $\gamma \ge 0$ such that $E\left\{|w(X,T)c(X,T)|^{2+\gamma}|X=x\right\}<\infty$.
\end{itemize}}

To describe the asymptotic normality of the spline estimator $\hat{\tau}(x)$ under the $L_1$ loss, we write matrix $\mathbf{D}$ with the $(i,j)$-th element being
\begin{equation*}
\begin{split}
\mathbf{D}_{ij}=&\int_0^1 \Big[p(x)w(x,1)c^2(x,1)f_1(g(x)+c(x,1)\tau_0(x)|x)\\
&+(1-p(x))w(x,-1)^2c(x,-1)f_{-1}(g(x)+c(x,-1)\tau_0(x)|x)\Big] B_{i}(x)B_{j}(x) dQ(x),
\end{split}
\end{equation*}
where $f_{1}(y|x)$ and $f_{-1}(y|x)$ are the conditional density functions of $Y^{(1)}$ and $Y^{(-1)}$ given $X=x$, respectively. {We give the following theorem for the spline-based LAD regression:}

\begin{theorem}
With conditions {C1-C6 and C12}, as $n \rightarrow \infty$, we have
$\sqrt{n/K_n}(\hat{\tau}(x)-\tau_0(x)-b^a(x)) \stackrel{D}{\rightarrow} N(0,\Psi(x)),$
where
$\Psi(x)=lim_{n \rightarrow \infty} \frac{1}{4K_n}B(x)^T\mathbf{D}^{-1}\mathbf{G}\mathbf{D}^{-1}B(x).$
\end{theorem}

\noindent\emph{Remark 7.} For inference concerning $\tau_0(x)$, the variance of the estimator can be obtained by using resampling methods, as the asymptotic variance is difficult to work with. In a simulation experiment in Appendix B.2.2, we show that the bootstrap C.I. consistent with theoretical C.I..

\section{A Simulation Study}

We conducted an extensive simulation study to evaluate the finite sample performance of the proposed methods. {We considered a large number of parameter settings, including four different learners under two different loss functions: (1) a robust version of the modified covariate method with efficiency augmentation ($L_1$-MCM-EA), (2) a robust R-learner ($L_1$-RL), (3) a robust A-learner ($L_1$-AL), (4) an $L_2$-based MCM-EA, (5) an $L_2$-based RL, (6) an $L_2$-based AL, and (7) a robust Q-learner ($L_1$-QL), and (8) an $L_2$-based Q-learner ($L_2$-QL). The first six methods are under the umbrella of A-learning and they are covered by the general formulation in (\ref{general}). The last two are Q-learning methods, which are not the focus of the current paper; we included them only for comparison. The first three methods are what we recommend for situations with a significant number of outliers; Methods 4-6 are standard $L_2$-based learners.} 

{
We used the A-learner described by Lu and colleagues \supercite{lu2013variable}. The objective functions of the A-learning methods 3 and 6 shared the same structure, except for the loss function $\rho$
$$L_n(\boldsymbol{\beta})=\frac{1}{n} \sum_{i=1}^n \rho \bigg( Y_i-\mathbf{X}_i^T \hat{\boldsymbol{\gamma}}-\left[\frac{T_i+1}{2}-\hat{p}(\mathbf{X}_i)\right]B(\mathbf{X}_i)^T\boldsymbol{\beta}\bigg)+\Lambda_n(\boldsymbol{\beta}),$$
where $\boldsymbol{\gamma}$ and $p(\mathbf{x})$ are estimated in advance. We estimated $\boldsymbol{\gamma}$ by regressing $Y$ on $\mathbf{X}$ using a linear regression, and $p(\mathbf{x})$ by regressing $\frac{T+1}{2}$ on $\mathbf{X}$ using GBM. The objective functions of the Q-learning methods 7 and 8 shared the same structure
$$L_n(\boldsymbol{\gamma},\boldsymbol{\beta})=\frac{1}{n}\sum_{i=1}^n \rho\left(Y_i-B(\mathbf{X}_i)^T\boldsymbol{\gamma}-\frac{T_i}{2}B(\mathbf{X}_i)^T\boldsymbol{\beta}\right)+\Lambda_n(\boldsymbol{\gamma},\boldsymbol{\beta}),$$
where we used $L_1$ or $L_2$ loss function for $\rho$. Note that a difference between the A-learner and R-learner is the choice of the augmentation. For A-learner we used a linear function as suggested by Lu  \supercite{lu2013variable} to estimate $\mu(\mathbf{X}_i)$; we used $L_1$-based GBM to estimate $\mu(\mathbf{X}_i)$ in the R-learner.

We designed the simulation study to assess the robustness of the $L_1$ and $L_2$-based methods, and to contrast the performance of the A and Q-learners.  We also examined the performance of the methods under different sample sizes, dimensionality, and proportions of outliers. }

{We assessed the performance of the methods using the standard metrics, including bias, variance, mean square error as well as mean absolute error. In addition, we  compared the value function $Q(\hat\eta)=E(Y(\hat\eta))$, i.e., the expected \emph{average} outcome under treatment $\hat\eta$, where $\hat\eta(\mathbf{x})=2I(\hat\tau(\mathbf{x})>0)-1$, as recommended by each method \supercite{zhang2012robust}. To estimate the $Q(\hat\eta)$ for a given regimen, we conducted a Monte Carlo simulation using model $Y(\hat\eta)=b_0(\mathbf{X})+\frac{\hat\eta}{2}\tau_0(\mathbf{X})+\varepsilon$, replacing $T$ in (\ref{model}) by $\hat\eta$, and we set the number of replicates is $10^6$. The value function calculated based on the true treatment effects was $E[Y(\eta^{opt})]=1.25$, where $\eta^{opt}(\mathbf{x})=2I(\tau_0(\mathbf{x})>0)-1$}.  We also assessed the sensitivity and specificity for variable selection under our penalty. With the number of simulation replication $R$, we defined
$$\begin{aligned}
MAE_v &=\frac{1}{R}\sum_{r=1}^{R} |\hat{\tau}^{(r)}(\mathbf{x}_v)-\tau_0(\mathbf{x}_v)|, \quad MSE_v =\frac{1}{R}\sum_{r=1}^{R} [\hat{\tau}^{(r)}(\mathbf{x}_v)-\tau_0(\mathbf{x}_v)]^2,\\
|Bias_v|^2 &= |\frac{1}{R}\sum_{r=1}^{R} \hat{\tau}^{(r)}(\mathbf{x}_v)-\tau_0(\mathbf{x}_v)|^2,\quad Var_v =\frac{1}{R}\sum_{r=1}^{R} [\hat{\tau}^{(r)}(\mathbf{x}_v)-\overline{\hat{\tau}(\mathbf{x}_v)}]^2\\
Sensitivity &=\frac{TP}{TP+FN}, \quad Specificity=\frac{TN}{TN+FP},\\
\end{aligned}$$
where $\mathbf{x}_v$ is the $v$-th observation from the validation set, $\hat{\tau}^{(r)}(\mathbf{x})$ is the estimator of $\tau(\mathbf{x})$ based on the $r$-th data replication, and $\overline{\hat{\tau}(\mathbf{x}_v)}$ is the average of all estimators of the  $v$-th observation. TP, FN, TN, and FP represented the numbers of true positive, false negative, true negative, and false positive. In this research, the size of the validation set $n_v$ was set to 200; we summarized the performance over the whole validation set by taking the averages (i.e.,$\overline{MSE}=\frac{1}{n_v}\sum_{v=1}^{n_v} MSE_v$). For simplicity, we reported $MSE$, $MAE$, $|Bias|^2$, and $Var$.

\subsection{Data Generation}

We generated data as follows, the dimension of the covariates was indexed by $p$: 
$$\begin{aligned}
&\mathbf{X}_i \sim N_p(0,\mathbf{\Sigma}),  diag(\mathbf{\Sigma})=\mathbf{1}, Corr(X_{ij},X_{ik})=0.5^{|j-k|}, i=1,...,n,\\
&D_i|\mathbf{X}_i \sim Bernoulli(p(\mathbf{X}_i)), T_i = 2D_i-1, logit(p(\mathbf{X}_i))=X_{i1}-X_{i2},\\
&Y_i = b_0(\mathbf{X}_i) + \frac{T_i}{2} \tau_0(\mathbf{X}_i)+\varepsilon_i,\varepsilon_i \sim (1-\xi_o)N(0,1)+\xi_o Laplace(0,10), \\
&b_0(\mathbf{X}_i)= 0.5+4 X_{i1} +X_{i2}-3X_{i3},
\tau_0(\mathbf{X}_i)=2sin(2X_{i1})-X_{i2}+3tanh(0.5X_{i3}),
\end{aligned}$$
where $\eta_o$ represented the proportion of outliers. We considered three settings: \emph{(1) Various levels of outliers} $\xi_o \in \{0,0.05,0.1,0.15,0.2\}$, with $n=1000$ and $p =10$; \emph{(2) Various training sample sizes} $n \in \{200, 500, 1000\}$, with $p=10$ and $\xi_o \in \{0,0.05\}$; \emph{(3) Various dimension of training sample} $p \in \{10, 30, 50\}$, with $n=1000$ and $\xi_o \in \{0,0.05\}$.

\subsection{Simulation Results}
\begin{figure}[H]
\centering
\includegraphics[scale=0.45]{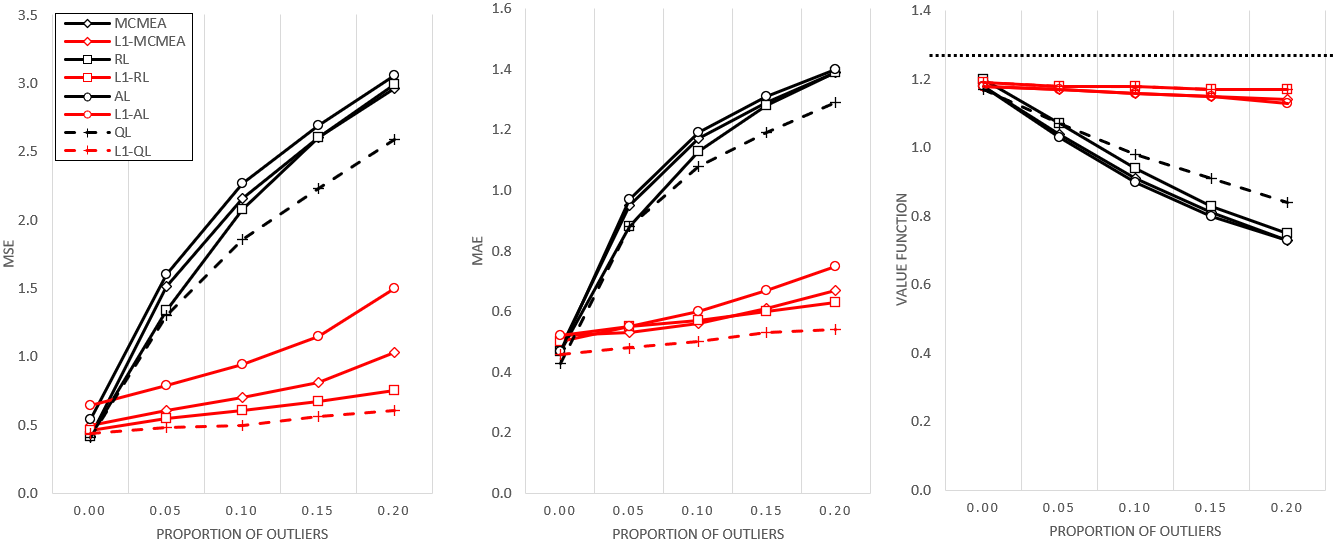}
\begin{minipage}[t]{0.9\textwidth} 
\caption{\footnotesize Comparison of mean squared error (MSE), mean absolute error (MAE), and value function ($Q(\hat{\eta})$) of the $L_1$-MCM-EA (red solid line), $L_1$-RL (red solid line), $L_1$-AL (red solid line), $L_1$-QL (red dashed line), MCM-EA (black solid line), RL (black solid line), AL (black solid line), and QL (black dashed line) under various levels of outliers. When there were outliers, both $L_1$-based methods outperformed the $L_2$-based methods. Advantage of the $L_1$-based methods increased with the proportion of outliers, under MSE, MAE, and $Q(\hat{\eta})$.}
\label{setup1}
\end{minipage}
\end{figure}

{Figure~\ref{setup1} showed that when there were outliers, the $L_1$-based methods uniformly outperformed the $L_2$-based methods under the MSE, MAE, and $Q(\hat{\eta})$ value. Advantage of the robust methods increased with the proportion of outliers. The robust R-learner outperformed the robust A-learner because $\mu(x)$ was not a linear function. And there were little practical differences between the robust R-learner and robust MCM-EA. The Q-learner performed the best under MSE and MAE because it is a one-step estimation procedure, and thus avoiding the errors associated with the nuisance quantity estimation. This is consistent  with the observations made by Schulte \supercite{schulte2014q} that the Q-learner tended to perform better than the standard A-learner \emph{when all models were correctly specified}. We conducted a separate simulation  for a setting where the Q-function was mis-specified. The results reported in Appendix B.2.3 showed that in the presence of outliers, bias in the mis-specified $L_1$-QL was larger than that of the $L_1$-MCMEA, $L_1$-RL, and $L_1$-AL. The same was also true for MSE and MAE. In terms of the value function $Q(\hat\eta)$, $L_1$-QL had smaller $Q(\hat\eta)$ values than methods under the A-learning umbrella; findings were consistent with MSE.} 

(Figure 1 goes here)

\begin{figure}[H]
\centering
\includegraphics[scale=0.5]{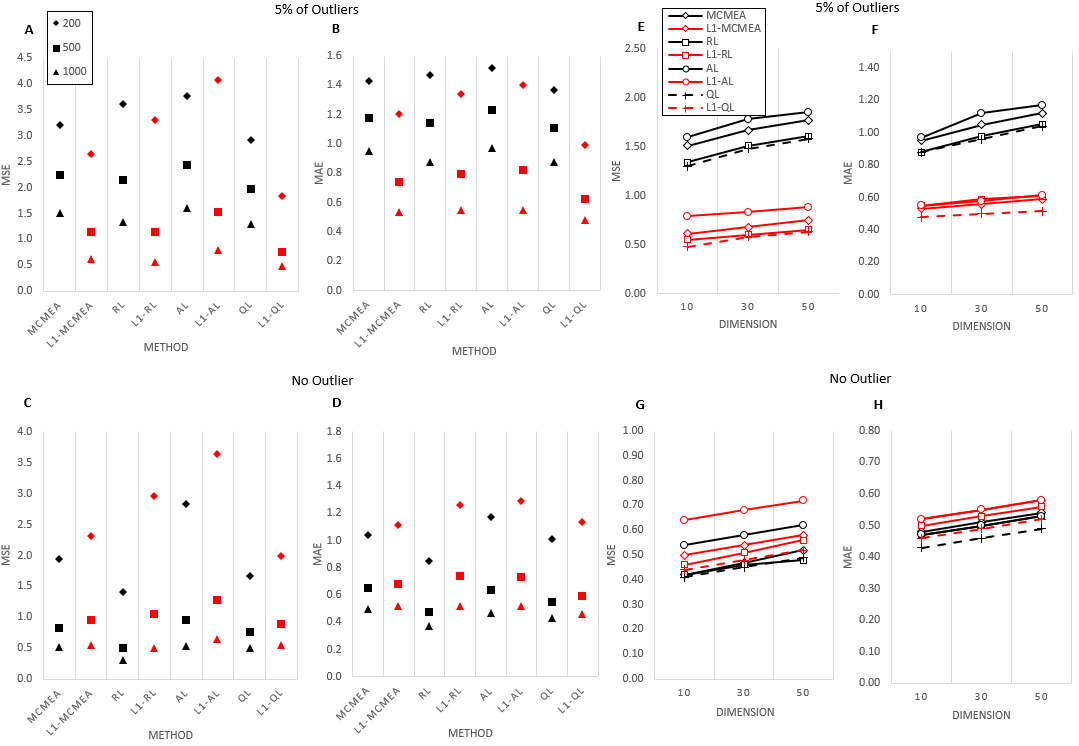}
\begin{minipage}[t]{0.9\textwidth} 
\caption{\footnotesize Panels A-D -- Mean squared error (MSE) and mean absolute error (MAE) values of different methods under different sample sizes, with and without outliers. The $L_1$-based methods are indicated by red symbols, whereas the $L_2$-based methods are indicated by black symbols. Panels E-H -- Impact of the dimension on different methods. $L_1$-based methods (red lines) are robust to outliers, whereas the $L_2$-based methods (black lines) are standard methods.}
\label{setup2}
\end{minipage}
\end{figure}

Figure~\ref{setup2} (A-D) showed the effects of sample size. Regardless of the presence or absence of outliers, as the sample size increased, MSE and MAE decreased for all methods. When there were no outliers, at a given sample size, the $L_2$-based methods tended to perform slightly better than the $L_1$-based methods, because the $L_2$-based methods were more efficient when the errors were normally distributed. But when there were even a small proportion of outliers, only 5\% of errors generated from a different distribution, the robust methods outperformed $L_2$-based methods by a noticeable margin. Figure~2 (E-H) showed that the performance of proposed methods without NIS did not change substantially as the dimension of the covariates increased.

(Figure 2 goes here)

Additional simulation details, including the squared bias, variance, MSE, MAE, sensitivity, specificity, and value function of the eight methods were reported in Appendix B.1. We have also examined the effects of dimension and smoothing on treatment effect estimation. Those results are included in Appendix B.2. 

We conducted additional simulation in one covariate setting, where we calculated the pointwise bootstrap confidence intervals for $\tau(x)$, under both $L_1$ and $L_2$ versions of the MCM-EA and RL methods, with and without penalty. The $L_1$-based methods generally produced coverage probabilities very close to the nominal level, even with the presence of outliers, whereas the $L_2$-based methods' coverage sometimes deviated strongly from $0.95$. See Appendix B.2, Table~B.5.

\section{Real Data Application}

To illustrate the methods we propose, we estimated the treatment effects of two different antihypertensive therapies by analyzing the observed clinical data set from the Indiana Network of Patient Care, a local EHR system. The data were a subset of a previous study assessing the blood pressure (BP)-lowering effects of various antihypertensive agents \supercite{tu2016triamterene}. This analysis compared the BP effects of angiotensin-converting-enzyme inhibitors (ACEI) alone and a combination of ACEI and hydrochlorothiazide (HCTZ). We considered those on ACEI alone as in treatment group A, and those on ACEI+HCTZ as in group B. The primary outcome of interest is clinically recorded systolic BP in response to these therapies. Independent variables included the demographic and clinical characteristics, as well as medication-use behaviors of the study participants. Data from $882$ participants were used in the current analysis. Among these, $350$ were on the monotherapy of ACEI, and $532$ were on the combination therapy of ACEI+HCTZ. Characteristics of the study participants are presented in Table~\ref{demographic}. {There were four continuous variables (pulse, BMI, age, and medication adherence) and 12 binary variables (gender, race, and ten comorbidities). The continuous variables were standardized before the analysis and expressed as linear combinations of splines.}

{We expressed the treatment effect of treatment B, in comparison against treatment A, as a function of the patient characteristics $\mathbf{x}$
$$\tau_0(\mathbf{x})=E[Y^{(B)}-Y^{(A)}|\mathbf{X}=\mathbf{x}],$$
where $Y^{(A)}, Y^{(B)}$ represented the potential systolic BP of ACEI alone group and ACEI+HCTA group.} Since the antihypertensive effect of a therapy is measured by its ability to lower BP, a negative $\tau(\mathbf{x})$ indicates a superior effect of the combination therapy over the monotherapy, for a given $\mathbf{x}$. An important covariate of interest was the level of medication adherence, which we measured with the proportions of days covered (PDC) by the medication.

Preliminary data examination showed that the observed systolic BP was right-skewed in both groups. The Shapiro–Wilk's test further confirmed that the systolic BP was not normally distributed, and there were outliers in the observed outcome (ACEI alone: $W=0.9912$, $p=0.035$; ACEI+HCTZ: $W=0.9617$, $p=1.498e-10$). We, therefore, used the $L_1$-based methods with additive B-splines to analyze the data. Here the B-splines were used to accommodate the possible nonlinear influences of the independent variables on the treatment effect.

Naive comparison of the systolic BP-effects between the two treatment strategies suggested that the combination therapy (ACEI+HCTZ) was significantly worse than the monotherapy (ACEI alone) in its ability to lower systolic BP (Table~1, $134.86$mm Hg in ACEI vs. $137.49$ mm Hg in ACEI+HCTZ; $p=0.004$). A similar difference was seen in diastolic BP ($80.98$mm Hg in ACEI vs. $82.26$ mm Hg in ACEI+HCTZ; $p=0.046$). The observation is counterintuitive because there are no known mechanisms that would explain the attenuated BP benefit of ACEI when HCTZ is added to the treatment regimen. In fact, the current clinical guidelines recommend HCTZ as the first-line therapy for essential hypertension \supercite{james20142014}. BP is regulated by hormones in the renin-angiotensin-aldosterone system (RAAS) \supercite{tu2017varying}. ACE inhibitors block the conversion of angiotensin I to angiotensin II, diminishing the latter's effects on aldosterone production and sodium retention and causing BP reduction. Thiazide diuretics lower BP by suppressing the extracellular fluid volume, which in turn reduces aldosterone secretion. Together, the two drugs are expected to have additive effects in lowering BP. In clinical practice, the two are often used concurrently.

(Table 1 goes here)

A closer examination of the characteristics of the patients on these therapies showed that patients on the combination therapy were older, more likely to be female, and overweight. Using GBM described in Section~2.3, we examined the mean function of systolic BP $\hat{\mu}(\mathbf{x})$ and the propensity of a patient receiving the combination therapy $\hat{p}(\mathbf{x})$. The estimated propensity score distributions were clearly different for the two treatment groups, whereas the mean functions were similar. See Appendix C. More specifically, the histogram of mean functions overlapped, indicating no apparent differences between the mean systolic BP between the two treatment groups. The different propensity score distributions of the two groups clearly showed that non-random treatment assignment. The importance levels of the covariates from GBM and additional modeling details were summarized in Appendix C. The systematic differences in patient characteristics between the two treatment groups suggested that a naive comparison was not appropriate and should not be trusted.

\begin{figure}[H]
\centering
\includegraphics[scale=0.48]{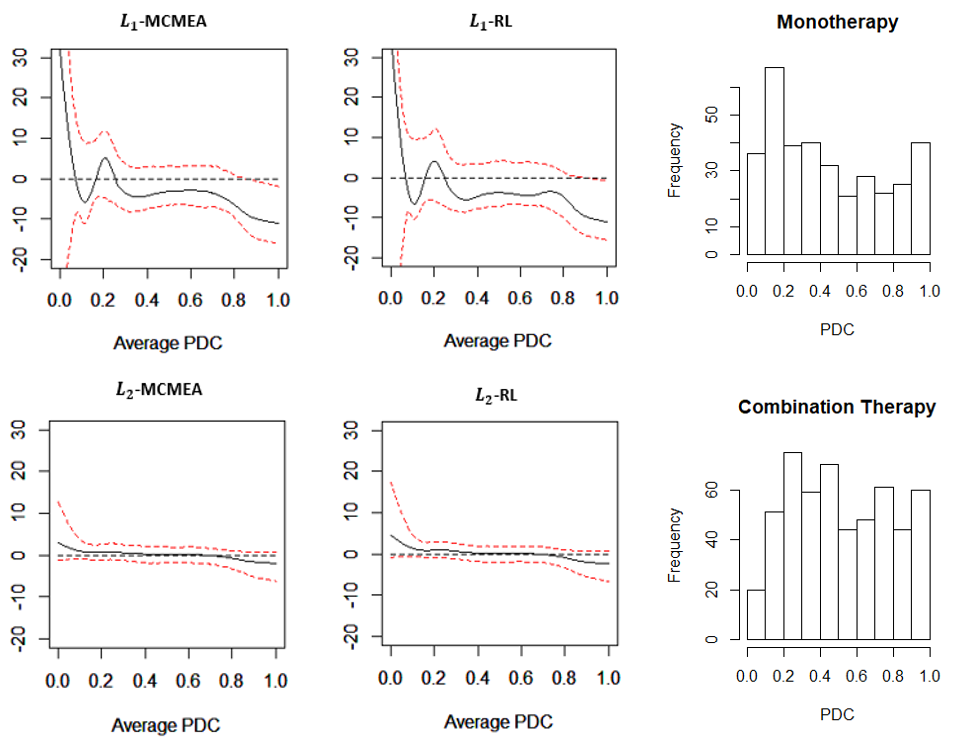}
\caption{\footnotesize Estimated treatment effects as functions of medication adherence (Proportion of Days Covered or PDC) under different methods. To plot these marginal effects, we fixed the continuous covariates at their mean values, and binary covariates at zero. There were 100 (11.3\%) patients that had PDC above $0.9$. Among these, 60 were in the combination therapy group, and 40 in the monotherapy group.}
\label{final_estimates}
\end{figure}

We then analyzed the data with the proposed methods. Importantly, both the $L_1$-MCM-EA and $L_1$-RL selected BMI and PDC in the final models. The $L_2$-based methods, on the other hand, only selected PDC. As we have shown in the simulation study, in the presence of outliers, the rates of correct selection of patient characteristics in the proposed methods were substantially greater than that of the $L_2$-based methods. The estimated treatment effects as functions of BMI and PDC were depicted in Figure~\ref{final_estimates}.

Figure~\ref{final_estimates} showed that $\hat{\tau}$ gradually decreased as the medication adherence measure PDC increased. Lower $\hat{\tau}$ indicated a stronger efficacy of the combination therapy than the monotherapy. Although decreasing trends were observed in both $L_1$ and $L_2$-based methods, the $L_2$ methods failed to detect any differences between the two therapies, as the 95\% confidence intervals for $\hat{\tau}(PDC)$ consistently covered zero. The $L_1$-based estimators, however, showed a superior blood pressure-lowering effect of the combination therapy, but only \emph{when PDC$>90\%$}. The fact that treatment effects varied with medication adherence should not be surprising.  As the former US Surgeon General, Dr. C. Everett Koop, wisely observed, ``Drugs don't work in patients who don't take them.'' \supercite{osterberg2005adherence} In this analysis, we do not expect significant differences between the treatments \emph{when patients are not adherent to the prescribed regimen.} Findings such as this are not unexpected in comparative effectiveness analysis of EHR data. Because unlike well-controlled clinical trials, few measures are in place to ensure patients faithfully take their medications in the real-world of clinical care. In the current application, the fact that the $L_1$-based estimators detected significant differences highlights the proposed methods' advantage. Using $L_1$-based estimators, we also examined the influences of BMI on $\tau$, which did not reach the level of statistical significance (data not shown).

(Figure 3 goes here)

To check the conditional independence error assumption, we performed the invariant residual distribution test (IRD-test), invariant environment prediction test (IEP-test), invariant conditional quantile prediction test (ICQP-test), invariant targeted prediction test (ITP-test) \supercite{heinze2018invariant}, and invariant residual prediction test (IRP-test) \supercite{shah2018goodness}. The conditional independence error assumption held for both proposed methods at the significant level of 0.05 (see Table~\ref{table:testresult}).

(Table 2 goes here)

{In addition to the marginal treatment effect, we also examined the value function $\hat{Q}(\hat\eta)$, which is the expected SBP under the estimated treatment regime $\hat\eta(\mathbf{x})=2I(\hat\tau(\mathbf{x})<0)-1$. In the absence of a true value function, we used a 10-fold cross validation to estimate $\hat{Q}(\hat\eta)$. For each fold $F_j$, we used the rest data for estimating $\hat{\mu}^{(-j)}(\mathbf{x})$, $\hat{p}^{(-j)}(\mathbf{x})$, and $\hat{\tau}^{(-j)}(\mathbf{x})$. Then we estimated the expected SBP by $\hat{Q}^{(j)}(\hat\eta) \triangleq \frac{1}{n_j} \sum_{i \in F_j} \hat{Y}^{(-j)}(\mathbf{x}_i)$ with $\hat{Y}^{(-j)}(\mathbf{x}_i)=\hat{\mu}^{(-j)}(\mathbf{x}_i)+[I(\hat{\tau}^{(-j)}(\mathbf{x}_i)<0)-\hat{p}^{(-j)}(\mathbf{x}_i)]\hat{\tau}^{(-j)}(\mathbf{x}_i)$. By looping over $j = 1,2,...,10$, we calculated $\hat{Q}(\hat{\eta})=\frac{1}{10}\sum_{j=1}^{10}\hat{Q}^{(j)}(\hat{\eta})$. The observed average SBP was $136.45$ mmHg, the estimates based on the $L_1$-MCMEA and $L_1$-RL were lower than the observed value. The estimates based on $L_2$-MCMEA and $L_2$-RL were slightly higher than the corresponding $L_1$-based methods. This results in Table \ref{table:value} showed that the SBP could be reduced if treatment were to be assigned in accordance with the  therapy recommended by the estimated treatment regime.} Table \ref{table:pdc} showed among the patients included in the analysis, 100 (11.3\%) had PDC above 90\%. We further examined the numbers of patients assigned to the two different treatment groups based on the estimated treatment effects. More patients would be assigned to the combination therapy group because it had a significantly greater blood pressure efficacy when patients take their medications. On the other hand, had we used the $L_2$ based methods, almost all of the patients would have been assigned to the monotherapy group, which contradicts the recommendations from the current clinical guidelines.

(Table 3 goes here)

(Table 4 goes here)

In summary, the naive and $L_2$-based methods showed that the combination therapy of ACEI and HCTZ had a worse BP-lowering effect than the monotherapy of ACEI, a finding that contradicts the recommendations of the current clinical guidelines of hypertension treatment. The $L_1$-based methods have produced results that are better explained by the existing clinical and biological evidence. The analysis showed that treatment effects tended to improve when patients adhere to their prescribed medications.

\section{Discussion}

{We started this work searching for a robust estimator for heterogeneous treatment effects that could be used in EHR analysis, where outliers often undermine the validity of estimation. In the process, we discovered a general formulation that not only addresses the issues of outliers but also covers a broad class of learners, including the commonly used A-learner, as well as other learning methods associated with it, such as the inverse propensity weighting, various modified outcome methods, modified covariate methods with or without efficiency augmentation, and the doubly robust method. } Through a clever specification of the weight and efficiency augmentation functions, the formulation not only brings together a diverse set of methods under a unified presentation but also facilitates the development of a general-purpose procedure for implementation. Although we have highlighted the use of the $L_1$ loss function for increased robustness against outliers in the EHR data, the score equation we described can readily accommodate other loss functions, giving the analyst much-enhanced flexibility in practical data analysis. As we have shown in our simulation studies, the use of $L_1$ loss function in heterogeneous treatment effect estimation substantially increases the estimation methods' robustness. Importantly, the gain in robustness does not appear to inflict a heavy toll on efficiency. Initial theoretical exploration suggests that reasonable asymptotic behavior can still be expected for the resultant estimators under various loss functions. Besides the flexibility in loss function selection, the general formulation also permits the incorporation of other useful features, such as nonparametric specifications of the mean and propensity functions and embedded dimensional reduction tools. 

{ A theoretical examination of the proposed method shows that the resultant estimators possess the desirable property of asymptotic normally, under fairly general regularity conditions, and various commonly used loss functions. }Simulation studies have provided strong and consistent empirical evidence on the utility of the proposed methods. Then through a real data application, we demonstrated how the proposed approach could be used in EHR data analysis to quantify treatment effects that varied with patient drug-taking behaviors. The findings are in line with the existing clinical understanding of the therapeutic effects of the treatments. This said, the proposed method's performance remains to be tested in a wider range of clinical applications. Notwithstanding this limitation, we have taken the first steps in developing a scalable solution to estimate heterogeneous treatment effects in settings that are more prone to various forms of data irregularities.  

\printbibliography

\newpage
\begin{table}[H]
\caption{Demographic and Clinical Characteristics of Study Subjects}
\label{demographic}
\begin{center}
\begin{tabular}{ |l|l|l|c| } 
\hline
Variable & ACEI (n=350) & ACEI+HCTZ (n=532) & p-value \\
\hline
 & \multicolumn{3}{c|}{mean (sd)} \\ 
\hline
Average Systolic BP & 134.86 (11.72) & 137.49 (14.11) & 0.004* \\ 
Average Diastolic BP & 80.98 (8.64) & 82.26 (9.77) & 0.046*\\
Pulse & 83.67 (10.36) & 81.12 (10.51) & <0.001*\\
BMI & 31.75 (8.65) & 33.39 (8.79) & 0.007*\\
Age & 47.83 (12.84) & 50.03 (12.43) & 0.012*\\
Medication Adherence (PDC) & 0.45 (0.30) & 0.52 (0.27) & <0.001*\\
\hline
&\multicolumn{3}{c|}{n (percentage)}\\
\hline
Male & 158 (45.1\%) & 189 (35.5\%) & 0.005*\\
Black & 144 (41.1\%) & 290 (54.5\%) & <0.001*\\
Diabetes & 155 (44.3\%) & 114 (21.4\%) & <0.001*\\
Chronic Kidney Disease (CKD) & 8 (2.3\%) & 13 (2.4\%) & 1.000\\
Coronary Artery Disease (CAD) & 10 (2.9\%) & 15 (2.8\%) & 1.000\\
Myocardial Infraction (MI) & 2 (0.6\%) & 3 (0.6\%) & 1.000\\
Congestive Heart Failure (CHF) & 7 (2.0\%) & 11 (2.1\%) & 1.000\\
Hyperlipidemia & 53 (15.1\%) & 88 (16.5\%) & 0.645\\
Atrial fibrillation & 1 (0.3\%) & 5 (0.9\%) & 0.461\\
Stroke & 9 (2.6\%) & 6 (1.1\%) & 0.175\\
Chronic Obstructive Pulmonary Disease (COPD) & 40 (11.4\%) & 51 (9.6\%) & 0.443\\
Depression & 88 (25.1\%) & 132 (24.8\%) & 0.975\\
\hline
\end{tabular}
\end{center}
\end{table}

\begin{table}[H]
\begin{center}\begin{threeparttable}
\caption{Conditional independence test results}
\label{table:testresult}
\begin{tabular}{ |l|c|c|c|c|c| } 
\hline
Method & IRD-test & IEP-test & ICQP-test & ITP-test & IRP-test \\
\hline
$L_1$-RL & 0.11 & 0.71 & 0.82 & 0.58 & 0.26 \\
$L_1$-MCM-EA & 0.25 & 0.71 & 1.00 & 0.57 & 0.20\\
\hline
\end{tabular}
\begin{tablenotes}
      \small
      \item Note: The values in the table are p-values. The conditional independence error assumption holds for both proposed methods at the significant level of 0.05.
\end{tablenotes}
\end{threeparttable}\end{center}
\end{table}

\begin{table}[H]
\caption{Value functions of methods considered in application}
\label{table:value}
\begin{center}\begin{threeparttable}
\begin{tabular}{ |l|p{2.5cm}|p{2.5cm}|p{2.5cm}|p{2.5cm}| } 
\hline
Method & $L_1$-RL & $L_2$-RL & $L_1$-MCM-EA & $L_2$-MCM-EA  \\
\hline
$\hat{Q}(\hat{\eta})$& 134.98 &  135.13 & 133.96 & 134.64 \\
\hline
\end{tabular}
\begin{tablenotes}
      \small
      \item Note: The value function is the expected systolic blood pressure under the estimated treatment regimen $\hat\eta(\mathbf{x})=2I(\hat\tau(\mathbf{x})<0)-1$. Differences in value functions of the $L_1$ and $L_2$-based methods are minimal. However, the $L_1$-based methods outperform the $L_2$-based methods  when data irregularities are present. See results from Table~\ref{table:pdc} and Figure~\ref{final_estimates}.
\end{tablenotes}
\end{threeparttable}\end{center}
\end{table}

\begin{table}[H]
\caption{Treatment Assignment of Observations with PDC>0.9}
\label{table:pdc}
\begin{center}\begin{threeparttable}
\begin{tabular}{ |l|c|c| } 
\hline
Method & Monotherapy (n) & Combination Therapy (n) \\
\hline
Real data & 40 & 60  \\
$L_1$-RL & 4 & 96  \\
$L_1$-MCM-EA & 2 & 98 \\
$L_2$-RL & 100 & 0  \\
$L_2$-MCM-EA & 98 & 2 \\
\hline
\end{tabular}
\begin{tablenotes}
      \small
      \item Note: Patients in the application data set whose $PDC>0.9$ are reassigned treatments by estimated treatment effect, i.e. $\hat\eta(\mathbf{x})=2I(\hat\tau(\mathbf{x})<0)-1$. Under the $L_1$-based methods, most of the patients will be assigned to the combination therapy group, consistent with the results in Figure~\ref{final_estimates}. However, under the $L_2$-based methods, most of the patients will be assigned to the monotherapy group, which is counter-intuitive, because when patient adhere to prescription, the combination therapy is known to be more efficacious.  
\end{tablenotes}
\end{threeparttable}\end{center}
\end{table}




\newpage

\newpage

\newpage

\newpage
\section*{Appendices}
\subsection*{A. Derivation and Proofs}
\subsubsection*{A.1. Expressing the existing methods in the general formulation}
\setcounter{equation}{0}
In A.1,  we specify the expressions of $c(\mathbf{X},\mathbf{T})$, $w(\mathbf{X},\mathbf{T})$, and $g(\mathbf{X})$ for MCM-EA, RL, IPW, and AIPW methods. We show they satisfy the constraints associated with the general formulation. For most of the methods, the derivations are similar for $L_1$ and $L_2$ loss functions. So we show the derivation under the $L_2$ loss.

\noindent\emph{(1) MCM-EA.} The objective function of $L_2$-MCM-EA method is
$$L(\tau(\mathbf{x})) = E\left[\frac{(Y_i-\mu(\mathbf{X}_i)-\frac{T_i}{2}\tau(\mathbf{X}_i))^2}{T_i p(\mathbf{X}_i)+(1-T_i)/2}\bigg|\mathbf{X}_i=\mathbf{x}\right].$$
We write
$$
w(\mathbf{X}_i,T_i)=\frac{1}{T_ip(\mathbf{X}_i)+(1-T_i)/2},\quad c(\mathbf{X}_i,T_i)= \frac{T_i}{2}, \quad g(\mathbf{X}_i)=\mu(\mathbf{X}_i).
$$
Then 
$$p(\mathbf{x})w(\mathbf{x},1)c(\mathbf{x},1)+(1-p(\mathbf{x}))w(\mathbf{x},-1)c(\mathbf{x},-1)=p(\mathbf{x})\frac{1}{p(\mathbf{x})}\frac{1}{2}+(1-p(\mathbf{x}))\frac{1}{1-p(\mathbf{x})}\left(-\frac{1}{2}\right)=0$$
$$c(\mathbf{x},1)-c(\mathbf{x},-1)=\frac{1}{2}-\left(-\frac{1}{2}\right)=1,$$
which shows the $c$ and $w$ functions satisfy Conditions C1 and C2. Condition C3 ($w>0$ and $c\ne 0$) is clearly met. The same set of parameters can be used in $L_1$ loss. The verification is the same.

\noindent\emph{(2) R-Learning.} The objective function of $L_2$-based R-learning method is
$$L(\tau(\mathbf{x})) = E\left[(Y_i-\mu(\mathbf{X}_i)-\frac{T_i-2p(\mathbf{X}_i)+1}{2}\tau(\mathbf{X}_i))^2\bigg|\mathbf{X}_i=\mathbf{x}\right].$$
We write 
$$
w(\mathbf{X}_i,T_i)=1,\quad c(\mathbf{X}_i,T_i)= \frac{T_i-2p(\mathbf{X}_i)+1}{2},\quad g(\mathbf{X}_i)=\mu(\mathbf{X}_i).
$$
Then 
$$p(\mathbf{x})w(\mathbf{x},1)c(\mathbf{x},1)+(1-p(\mathbf{x}))w(\mathbf{x},-1)c(\mathbf{x},-1)=p(\mathbf{x})(1-p(\mathbf{x}))+(1-p(\mathbf{x}))(-p(\mathbf{x}))=0$$
$$c(\mathbf{x},1)-c(\mathbf{x},-1)=(1-p(\mathbf{x}))-(-p(\mathbf{x}))=1.$$
Therefore, Conditions C1-C3 are met. The same specification works for $L_1$ loss. The verification of A-learning remains the same.

\noindent\emph{(3) IPW}. The objective function of $L_2$-based IPW method is
$$L(\tau(\mathbf{x})) = E\left[\left(\left(\frac{T_i+1}{2p(\mathbf{X}_i)}-\frac{1-T_i}{2(1-p(\mathbf{X}_i))}\right)Y_i-\tau(\mathbf{X}_i)\right)^2\bigg|\mathbf{\mathbf{X}_i}=\mathbf{x}\right].$$
We write 
$$
w(\mathbf{X}_i,T_i)=\left(\frac{T_i+1}{2p(\mathbf{X}_i)}-\frac{1-T_i}{2(1-p(\mathbf{X}_i))}\right)^2,\quad c(\mathbf{X}_i,T_i)= \frac{2p(\mathbf{X}_i)(1-p(\mathbf{X}_i))}{T_i-2p(\mathbf{X}_i)+1},\quad g(\mathbf{X}_i)=0.
$$
Then
$$p(\mathbf{x})w(\mathbf{x},1)c(\mathbf{x},1)+(1-p(\mathbf{x}))w(\mathbf{x},-1)c(\mathbf{x},-1)=p(\mathbf{x})\frac{1}{p(\mathbf{x})^2}p(\mathbf{x})+(1-p(\mathbf{x}))\frac{1}{(1-p(\mathbf{x}))^2}(p(\mathbf{x})-1)=0$$
$$c(\mathbf{x},1)-c(\mathbf{x},-1)=p(\mathbf{x})-(p(\mathbf{x})-1)=1.$$
Therefore, Conditions C1-C3 are met. The same specification works for the $L_1$ loss function. 

\noindent\emph{(4) AIPW.} The verification of AIPW method is the same.

\subsubsection*{A.2.  Basic properties of the general formulation}
\textbf{Property 1.} Under conditions C1-C3, $\tau_0(\mathbf{x})=argmin_{\tau(\mathbf{x})}E[w(\mathbf{X}_i,T_i)(y-g(\mathbf{X}_i)-c(\mathbf{X}_i,T_i)\tau(\mathbf{x}))^2|\mathbf{X}_i=\mathbf{x}]$.

\noindent\textit{Proof of Property 1.} 
$$\begin{aligned}
L(\tau(\mathbf{x}))=& E[w(\mathbf{X}_i,T_i)(Y_i-g(\mathbf{X}_i)-c(\mathbf{X}_i,T_i)\tau(\mathbf{X}_i))^2|\mathbf{X}_i=\mathbf{x}]\\
=& p(\mathbf{x})E[w(\mathbf{X}_i,T_i)(Y_i-g(\mathbf{X}_i)-c(\mathbf{X}_i,T_i)\tau(\mathbf{X}_i))^2|\mathbf{X}_i=\mathbf{x},T_i=1]\\
&+(1-p(\mathbf{x}))E[w(\mathbf{X}_i,T_i)(Y_i-g(\mathbf{X}_i)-c(\mathbf{X}_i,T_i)\tau(\mathbf{X}_i))^2|\mathbf{X}_i=\mathbf{x},T_i=-1]\\
=& p(\mathbf{x})w(\mathbf{x},1)E[(Y_i-g(\mathbf{X}_i)-c(\mathbf{X}_i,T_i)\tau(\mathbf{X}_i))^2|\mathbf{X}_i=\mathbf{x},T_i=1]\\
&+(1-p(\mathbf{x}))w(\mathbf{x},-1)E[(Y_i-g(\mathbf{X}_i)-c(\mathbf{X}_i,T_i)\tau(\mathbf{X}_i))^2|\mathbf{X}_i=\mathbf{x},T_i=-1]\\
\frac{\partial L(\tau(\mathbf{x}))}{\partial \tau(\mathbf{x})}=&   -2p(\mathbf{x})w(\mathbf{x},1)c(\mathbf{x},1) (E[Y_i^{(1)}|\mathbf{X}_i=\mathbf{x}]-g(\mathbf{X}_i)-c(\mathbf{x},1)\tau(\mathbf{x}))\\
&- 2(1-p(\mathbf{x}))w(\mathbf{x},-1)c(\mathbf{x},-1) (E[Y_i^{(-1)}|X_i=\mathbf{x}]-g(\mathbf{x})-c(\mathbf{x},-1)\tau(\mathbf{x}))\\
=&-2p(\mathbf{x})w(\mathbf{x},1)c(\mathbf{x},1) (b_0(\mathbf{x})+\frac{\tau_0(\mathbf{x})}{2}+E[\varepsilon_i^{(1)}|X_i=\mathbf{x}]-g(\mathbf{X}_i)-c(\mathbf{x},1)\tau(\mathbf{x}))\\
&- 2(1-p(\mathbf{x}))w(\mathbf{x},-1)c(\mathbf{x},-1) (b_0(\mathbf{x})-\frac{\tau_0(\mathbf{x})}{2}+E[\varepsilon_i^{(-1)}|X_i=\mathbf{x}]-g(\mathbf{x})-c(\mathbf{x},-1)\tau(\mathbf{x}))\\
\end{aligned}$$
Conditions C1-C3 and the conditional independence assumption lead us to $\tau_0(\mathbf{x})=argmin_{\tau(x)}L(\tau(\mathbf{x}))$. $\blacksquare$

\noindent\textbf{Property 2.} When $c(\mathbf{x},1)=1-p(\mathbf{x})$, the optimal augmentation function is the mean outcome function, i.e., $g_0(\mathbf{x})=\mu(\mathbf{x})$.

\noindent\textit{Proof of Property 2.}
We provide the optimal $g(\cdot)$ in this section, by optimality we mean the $g(\cdot)$ that minimizes the variance of estimator. Let $S(Y_i,\mathbf{X}_i,T_i;\tau(\mathbf{X}_i))$ be the derivative of the objective function $ w(\mathbf{X}_i,T_i)(Y_i-g(\mathbf{X}_i)-c(\mathbf{X}_i,T_i)\tau(\mathbf{X}_i))^2$, with respect to $\tau$. Then the estimating equation is
$$\begin{aligned}
\frac{1}{n}\sum_{i=1}^n S(Y_i,\mathbf{X}_i,T_i;\tau(\mathbf{X}_i))=&\frac{1}{n}\sum_{i=1}^n -2w(\mathbf{X}_i,T_i)c(\mathbf{X}_i,T_i)(Y_i-g(\mathbf{X}_i)-c(\mathbf{X}_i,T_i)\tau(\mathbf{X}_i))\\
=& \frac{1}{n}\sum_{i=1}^n S_0(Y_i,\mathbf{X}_i,T_i;\tau(\mathbf{X}_i))+2w(\mathbf{X}_i,T_i)c(\mathbf{X}_i,T_i)g(\mathbf{X}_i)=0,
\end{aligned}$$
where $S_0(Y_i,\mathbf{X}_i,T_i;\tau(\mathbf{X}_i))=-2w(\mathbf{X}_i,T_i)c(\mathbf{X}_i,T_i)[Y_i-c(\mathbf{X}_i,T_i)\tau(\mathbf{X}_i)]$ is the score function without augmentation. By Condition C1,
$E[2w(\mathbf{X}_i,T_i)c(\mathbf{X}_i,T_i)g(\mathbf{X}_i)]=0,$
the solution of the augmented score equation always converges to $\tau_0(\cdot)$ in probability. Following \supercite{tian2014simple,chen2017general}, selecting the optimal $g(\cdot)$ is equivalent to minimizing the conditional variance of 
$$\begin{aligned}
S_0(Y_i,\mathbf{X}_i,T_i;\tau_0(\mathbf{X}_i))+2w(\mathbf{X}_i,T_i)c(\mathbf{X}_i,T_i)g(\mathbf{X}_i),
\end{aligned}$$
where $\tau_0(\mathbf{x})$ is the minimizer of $E[w(\mathbf{X}_i,T_i)(Y_i-c(\mathbf{X}_i,T_i)\tau(\mathbf{X}_i))^2|\mathbf{X}_i=\mathbf{x}]$. Noting that
$$\begin{aligned}
&E[\{S_0(Y_i,\mathbf{X}_i,T_i;\tau_0(\mathbf{X}_i))+2w(\mathbf{X}_i,T_i)c(\mathbf{X}_i,T_i)g(\mathbf{X}_i)\}^2|\mathbf{X}_i=\mathbf{x}]\\
=&E[\{S_0(Y_i,\mathbf{X}_i,T_i;\tau_0(\mathbf{X}_i))+2w(\mathbf{X}_i,T_i)c(\mathbf{X}_i,T_i)g_0(\mathbf{X}_i)\}^2|\mathbf{X}_i=\mathbf{x}]\\
&+E[\{2w(\mathbf{X}_i,T_i)c(\mathbf{X}_i,T_i)(g_0(\mathbf{X}_i)-g(\mathbf{X}_i))\}^2|\mathbf{X}_i=\mathbf{x}]\\
\ge & E[\{S_0(Y_i,\mathbf{X}_i,T_i;\tau_0(\mathbf{X}_i))+2w(\mathbf{X}_i,T_i)c(\mathbf{X}_i,T_i)g_0(\mathbf{X}_i)\}^2|\mathbf{X}_i=\mathbf{x}],
\end{aligned}$$
where 
$g_0(\mathbf{x})=(1-p(\mathbf{x}))E[Y_i^{(1)}-c(\mathbf{X}_i,T_i)\tau(\mathbf{X}_i)|\mathbf{X}_i=\mathbf{x},T_i=1]+p(\mathbf{x})E[Y_i^{(-1)}-c(\mathbf{X}_i,T_i)\tau(\mathbf{X}_i)|\mathbf{X}_i=\mathbf{x},T_i=-1],$
which satisfies the equation
$$E[\{S_0(Y_i,\mathbf{X}_i,T_i;\tau_0(\mathbf{X}_i))+2w(\mathbf{X}_i,T_i)c(\mathbf{X}_i,T_i)g_0(\mathbf{X}_i)\}2w(\mathbf{X}_i,T_i)c(\mathbf{X}_i,T_i)\eta(\mathbf{X}_i)|\mathbf{X}_i=\mathbf{x}]=0$$
for any function $\eta(\cdot)$. By interaction model (1) and Condition C2, the expression of $g_0(\mathbf{x})$ can be further simplified to 
$g_0(\mathbf{x})=\mu(\mathbf{x})+[1-p(\mathbf{x})-c(\mathbf{x},1)]\tau_0(\mathbf{x}).$
As $\tau_0(\cdot)$ is the unknown target, when $c(\mathbf{x},1)=1-p(\mathbf{x})$, the optimal augmentation function is mean outcome function, i.e., $g_0(\mathbf{x})=\mu(\mathbf{x})$. $\blacksquare$

\noindent\textbf{Property 3.} Under Conditions C1-C3, $\tau_0(\mathbf{x})=argmin_{\tau(\mathbf{x})}E[w(\mathbf{X}_i,T_i)|y-g(\mathbf{X}_i)-c(\mathbf{X}_i,T_i)\tau(\mathbf{x}))||\mathbf{X}_i=\mathbf{x},T_i=t]$.

\noindent\textit{Proof of Property 3.}

$$\begin{aligned}
L(\tau(\mathbf{x}))=& E[w(\mathbf{X}_i,T_i)|Y_i-g(\mathbf{X}_i)-c(\mathbf{X}_i,T_i)\tau(\mathbf{X}_i)||\mathbf{X}_i=\mathbf{x},T_i=t]\\
=& p(\mathbf{x})E[w(\mathbf{X}_i,T_i)|Y_i-g(\mathbf{X}_i)-c(\mathbf{X}_i,T_i)\tau(\mathbf{X}_i)||\mathbf{X}_i=\mathbf{x},T_i=1]\\
&+(1-p(\mathbf{x}))E[w(\mathbf{X}_i,T_i)|Y_i-g(\mathbf{X}_i)-c(\mathbf{X}_i,T_i)\tau(\mathbf{X}_i)||\mathbf{X}_i=\mathbf{x},T_i=-1]\\
=& p(\mathbf{x})w(\mathbf{x},1)E[|Y_i-g(\mathbf{X}_i)-c(\mathbf{X}_i,T_i)\tau(\mathbf{X}_i)||\mathbf{X}_i=\mathbf{x},T_i=1]\\
&+(1-p(\mathbf{x}))w(\mathbf{x},-1)E[|Y_i-g(\mathbf{X}_i)-c(\mathbf{X}_i,T_i)\tau(\mathbf{X}_i)||\mathbf{X}_i=\mathbf{x},T_i=-1]
\end{aligned}$$

$$\begin{aligned}
\frac{\partial L(\tau(\mathbf{x}))}{\partial \tau(\mathbf{x})}=&   -p(\mathbf{x})w(\mathbf{x},1)c(\mathbf{x},1) E[sgn(Y_i-g(\mathbf{X}_i)-c(\mathbf{X}_i,T_i)\tau(\mathbf{X}_i))|\mathbf{X}_i=\mathbf{x},T_i=1]\\
&- (1-p(\mathbf{x}))w(\mathbf{x},-1)c(\mathbf{x},-1) E[sgn(Y_i-g(\mathbf{X}_i)-c(\mathbf{X}_i,T_i)\tau(\mathbf{X}_i))|\mathbf{X}_i=\mathbf{x},T_i=-1]\\
=&   -p(\mathbf{x})w(\mathbf{x},1)c(\mathbf{x},1) E[1-2I(Y_i-g(\mathbf{X}_i)-c(\mathbf{X}_i,T_i)\tau(\mathbf{X}_i))|\mathbf{X}_i=\mathbf{x},T_i=1]\\
&- (1-p(\mathbf{x}))w(\mathbf{x},-1)c(\mathbf{x},-1)E[1-2I(Y_i-g(\mathbf{X}_i)-c(\mathbf{X}_i,T_i)\tau(\mathbf{X}_i))|\mathbf{X}_i=\mathbf{x},T_i=-1]\\
=& -p(\mathbf{x})w(\mathbf{x},1)c(\mathbf{x},1)(1-2P(Y_i^{(1)}<g(\mathbf{X}_i)+c(\mathbf{X}_i,T_i)\tau(\mathbf{X}_i)|\mathbf{X}_i=\mathbf{x},T_i=1))\\
-& (1-p(\mathbf{x}))w(\mathbf{x},-1)c(\mathbf{x},-1)(1-2P(Y_i^{(-1)}<g(\mathbf{X}_i)+c(\mathbf{X}_i,T_i)\tau(\mathbf{X}_i)|\mathbf{X}_i=\mathbf{x},T_i=-1))\\
=& -p(\mathbf{x})w(\mathbf{x},1)c(\mathbf{x},1)(1-2F_{Y_i^{(1)}}(g(\mathbf{X}_i)+c(\mathbf{X}_i,T_i)\tau(\mathbf{X}_i)|\mathbf{X}_i=\mathbf{x},T_i=1))\\
&- (1-p(\mathbf{x}))w(\mathbf{x},-1)c(\mathbf{x},-1)(1-2F_{Y_i^{(-1)}}(g(\mathbf{X}_i)+c(\mathbf{X}_i,T_i)\tau(\mathbf{X}_i)|\mathbf{X}_i=\mathbf{x},T_i=-1))\\
\end{aligned}$$
By Condition C1, the score equation can be written to
$$
 F_{Y^{(1)}_i}(g(\mathbf{x})+c(\mathbf{x},1)\tau(\mathbf{x}))-F_{Y^{(-1)}_i}(g(\mathbf{x})+c(\mathbf{x},-1)\tau(\mathbf{x}))=0.
$$ 
Let $F_{Y^{(1)}_i}(g(\mathbf{x})+c(\mathbf{x},1)\hat{\tau}(\mathbf{x}))=F_{Y^{(-1)}_i}(g(\mathbf{x})+c(\mathbf{x},-1)\hat{\tau}(\mathbf{x}))=q$, where $q \in (0,1)$, then
$$\begin{aligned}
g(\mathbf{x})+c(\mathbf{x},1)\hat{\tau}(\mathbf{x})&=Q_q(Y^{(1)}_i|\mathbf{X}_i=\mathbf{x})\\
g(\mathbf{x})+c(\mathbf{x},-1)\hat{\tau}(\mathbf{x})&=Q_q(Y^{(-1)}_i|\mathbf{X}_i=\mathbf{x}).
\end{aligned}$$
By Condition C2 ($c(\mathbf{x},1)-c(\mathbf{x},-1)=1$), we have
$$\hat{\tau}(\mathbf{x})=Q_q(Y^{(1)}_i|\mathbf{X}_i=\mathbf{x})-Q_q(Y^{(-1)}_i|\mathbf{X}_i=\mathbf{x}).$$
As
$$\begin{aligned}
&Q_q(Y_i^{(1)}|\mathbf{X}_i=\mathbf{x})-Q_q(Y_i^{(-1)}|\mathbf{X}_i=\mathbf{x})\\
=&Q_q(Y_i|\mathbf{X}_i=\mathbf{x},T_i=1)-Q_q(Y_i|\mathbf{X}_i=\mathbf{x},T_i=-1)\\
=& Q_q(b_0(\mathbf{X}_i)+\frac{\tau_0(\mathbf{X}_i)}{2}+\varepsilon_i|\mathbf{X}_i=\mathbf{x},T_i=1)-Q_q(b_0(\mathbf{X}_i)-\frac{\tau_0(\mathbf{X}_i)}{2}+\varepsilon_i|\mathbf{X}_i=\mathbf{x},T_i=-1)\\
=& b_0(\mathbf{x})+\frac{\tau_0(\mathbf{x})}{2}+Q_q(\varepsilon_i|\mathbf{X}_i=\mathbf{x},T_i=1)-b_0(\mathbf{x})+\frac{\tau_0(\mathbf{x})}{2}-Q_q(\varepsilon_i|\mathbf{X}_i=\mathbf{x},T_i=-1)\\
=&\tau_0(\mathbf{x})+Q_q(\varepsilon_i|\mathbf{X}_i=\mathbf{x},T_i=1)-Q_q(\varepsilon_i|\mathbf{X}_i=\mathbf{x},T_i=-1).
\end{aligned}$$
By Assumption 3, $\tau_0(\mathbf{x})=argmin_{\tau(\mathbf{x})}L(\tau(\mathbf{x}))$. $\blacksquare$

\subsubsection*{A.3.  Asymptotic Properties}
To prove Theorem 1, we first introduce two lemmas.

\noindent\textbf{Lemma 1.} Under the same assumptions as Theorem 1, $\mathbf{W}_n$ is asymptotically equivalent to the $(K+p)$-dimensional normal with mean 0 and variance $\mathbf{G}$.

\noindent\textit{Proof of Lemma 1.} 
Let $Z_n=-\sqrt{\frac{K_n}{n}}\sum_{i=1}^n w(X_i,T_i)c(X_i,T_i)B(X_i)^T\boldsymbol\delta\rho'(U_i)$, the conditional expectation of $w(X_i,T_i)c(X_i,T_i)\rho'(U_i)$ with respect to $X_i$ is as follows. First, we calculate the conditional expectation with respect of $X_i$ and $T_i$,
\begin{align}
&E[w(X_i,T_i)c(X_i,T_i)\rho'(U_i)|X_i=x_i] \nonumber\\
=& E[w(X_i,T_i)c(X_i,T_i)\rho'(Y_i-g(X_i)-c(X_i,T_i)B(X_i)^T\boldsymbol\beta^*)|X_i=x_i] \nonumber\\
=& p(x_i)w(x_i,1)c(x_i,1)E[\rho'(Y_i^{(1)}-g(X_i)-c(X_i,1)B(X_i)^T\boldsymbol\beta^*)|X_i=x_i,T_i=1] \nonumber\\
&+(1-p(x_i))w(x_i,-1)c(x_i,-1)E[\rho'(Y_i^{(-1)}-g(X_i)-c(X_i,-1)B(X_i)^T\boldsymbol\beta^*)|X_i=x_i,T_i=-1] \\
=& p(x_i)w(x_i,1)c(x_i,1)\bigg(E[\rho'(Y_i^{(1)}-g(X_i)-c(X_i,1)B(X_i)^T\boldsymbol\beta^*) \nonumber\\
&-\rho'(Y_i^{(-1)}-g(X_i)-c(X_i,-1)B(X_i)^T\boldsymbol\beta^*)|X_i=x_i]\bigg).
\end{align}
From (1) to (2) is based on Condition C1. Then, based on the interaction model and the distance between $\tau_0(x)$ and $B(x)^T\boldsymbol\beta^*$, we have
\begin{align}
&E[w(X_i,T_i)c(X_i,T_i)\rho'(U_i)|X_i=x_i] \nonumber\\
=& p(x_i)w(x_i,1)c(x_i,1)\times \nonumber\\
&\bigg\{E[\rho'(b(X_i)+\frac{1}{2}\tau_0(X_i)+\varepsilon_i^{(1)}-g(X_i)-c(X_i,1)\tau_0(X_i)-c(X_i,1)b^a(X_i)[1+\Tilde{o}(1)]) \nonumber\\
&-\rho'(b(X_i)-\frac{1}{2}\tau_0(X_i)+\varepsilon_i^{(-1)}-g(X_i)-c(X_i,-1)\tau_0(X_i)-c(X_i,-1)b^a(X_i)[1+\Tilde{o}(1)])|X_i=x_i]\bigg\}\\
=& p(x_i)w(x_i,1)c(x_i,1)\times \nonumber\\
&\bigg\{E[\rho'(b(X_i)-g(X_i)-[c(X_i,1)-0.5]\tau_0(X_i)+\varepsilon_i^{(1)}-c(X_i,1)b^a(X_i)[1+\Tilde{o}(1)]) \nonumber\\
&-\rho'(b(X_i)-g(X_i)-[c(X_i,1)-0.5]\tau_0(X_i)+\varepsilon_i^{(-1)}-c(X_i,-1)b^a(X_i)[1+\Tilde{o}(1)])|X_i=x_i]\bigg\},
\end{align}
where $\Tilde{o}(1)$ uniformly holds for all $x$ by the distance between $\tau_0(x)$ and $B(x)^T\boldsymbol\beta^*$. From (3) to (4) is based on Condition C2. Let $\varphi(X_i,T_i)=b(X_i)-g(X_i)-[c(X_i,T_i)-0.5]\tau_0(X_i)+\varepsilon_i^{(T_i)}$, the expectation condition of $w(X_i,T_i)c(X_i,T_i)\rho'(U_i)$ on $X_i$ is

\begin{align}
& p(x_i)w(x_i,1)c(x_i,1)\times \bigg\{E[\rho'(\varphi(X_i,T_i))|X_i=x_i,T_i=1]-E[\rho'(\varphi(X_i,T_i))|X_i=x_i,T_i=-1] \nonumber\\
&-E[\rho''(\varphi(X_i,T_i)-\alpha^{(1)}c(X_i,1)b^a(X_i)[1+\Tilde{o}(1)])c(X_i,1)b^a(X_i)[1+\Tilde{o}(1)]|X_i=x_i]\nonumber\\
&+E[\rho''(\varphi(X_i,T_i)-\alpha^{(-1)}c(X_i,-1)b^a(X_i)[1+\Tilde{o}(1)])c(X_i,-1)b^a(X_i)[1+\Tilde{o}(1)]|X_i=x_i]\bigg\}\\
=& p(x_i)w(x_i,1)c(x_i,1)\times \nonumber\\
&\bigg\{-E[\rho''(\varphi(X_i,T_i)-\alpha^{(1)}c(X_i,1)b^a(X_i)[1+\Tilde{o}(1)])c(X_i,1)b^a(X_i)[1+\Tilde{o}(1)]|X_i=x_i] \nonumber\\
&+E[\rho''(\varphi(X_i,T_i)-\alpha^{(-1)}c(X_i,-1)b^a(X_i)[1+\Tilde{o}(1)])c(X_i,-1)b^a(X_i)[1+\Tilde{o}(1)]|X_i=x_i]\bigg\}, \nonumber
\end{align}
where $\alpha^{(1)},\alpha^{(-1)} \in (0,1)$ are from Taylor expansion. The first two terms in (5) inside the brace are cancelled out based on the conditional independence error assumption (Assumption 3). Finally, by the definition of $\Phi$, we have the conditional expectation equals
\begin{align}
& p(x_i)w(x_i,1)c(x_i,1)b^a(x_i)\times \nonumber\\
&\bigg\{-\Phi''([1-\alpha^{(1)}]c(X_i,1)b_a(X_i)[1+\Tilde{o}(1)]|X_i,T_i=1)c(x_i,1)) \nonumber\\
&+\Phi''([1-\alpha^{(-1)}]c(X_i,-1)b_a(X_i)[1+\Tilde{o}(1)]|X_i,T_i=-1)c(x_i,-1))\bigg\}[1+\Tilde{o}(1)] \nonumber\\
=& o(1). 
\end{align}
As the order of $b^a(x)$ is $o(K_n^{-(q+1)})$, Assumption 2 (positivity assumption), and Conditions C4 and C9, the conditional expectation is of $o(1)$. This means the conditional expectation of the score function with loss functions satisfy Conditions C7-C11 goes to zero.

Therefore, let $\psi(X_i,T_i,U_i)=w(X_i,T_i)c(X_i,T_i)\rho'(U_i)$, we obtain
$$\begin{aligned}
&E\left[\left|\sqrt{\frac{K_n}{n}}B(X_i)^T\boldsymbol\delta \left[\psi(X_i,T_i,U_i)-E[\psi(X_i,T_i,U_i)|X_i=x_i]\right]\right|^{2+\gamma}\bigg|X_i=x_i\right]\\
=& \left(\frac{K_n}{n}\right)^{\frac{2+\gamma}{2}} |B(x_i)^T\boldsymbol\delta|^{2+\gamma} E\left[\left|\psi(X_i,T_i,U_i)\right|^{2+\gamma}+o(1)\bigg|X_i=x_i\right]\\
=& \left(\frac{K_n}{n}\right)^{\frac{2+\gamma}{2}} |B(x_i)^T\boldsymbol\delta|^{2+\gamma} \{p(x_i)E[|w(x_i,1)c(x_i,1)\rho'(U_i)|^{2+\gamma}+o(1)|X_i=x_i,T_i=1]\\
&+(1-p(x_i))E[|w(x_i,-1)c(x_i,-1)\rho'(U_i)|^{2+\gamma}+o(1)|X_i=x_i,T_i=-1]\}\\
\le& O\left(\left(\frac{K_n}{n}\right)^{\frac{2+\gamma}{2}}\right),
\end{aligned}$$

where the last two steps are derived by Condition C11. The conditional variance of $Z_n$ respect to $X^{(n)}$ can be calculated as following.
\begin{align}
V[Z_n|X^{(n)}]&= \frac{K_n}{n} \sum_{i=1}^n \{B(x_i)^T\boldsymbol\delta\}^2 V\left[w(X_i,T_i)c(X_i,T_i)\rho'(U_i)\bigg|X_i=x_i\right]\nonumber\\
&= \frac{K_n}{n} \sum_{i=1}^n \{B(x_i)^T\boldsymbol\delta\}^2 \bigg\{E\left[\left(w(X_i,T_i)c(X_i,T_i)\rho'(U_i)\right)^2\bigg|X_i=x_i\right] \nonumber\\
&- E\left[w(X_i,T_i)c(X_i,T_i)\rho'(U_i)\bigg|X_i=x_i \right]^2 \bigg\}\\
&= K_n\boldsymbol\delta^T\mathbf{G}\boldsymbol\delta(1+o_P(1)) \\
&= O(K_n) \nonumber,
\end{align}
where $\mathbf{G}$ is the variance of $\mathbf{W}_n$. Here the derivation from (7) to (8) uses the Condition C8. Because the matrix $\mathbf{G}$ is positive definite and has a finite maximum eigenvalue for any bounded function (Lemma 6.2 of \supercite{zhou1998local}), there exists the constants $d_1$ and $d_2$ such that
$$d_1 \le \boldsymbol\delta^T \mathbf{G} \boldsymbol\delta \le d_2.$$
So it follows that
\begin{align*}
    \frac{1}{V[Z_n|X^{(n)}]^{(2+\gamma)/2}}\sum_{i=1}^n &E\left[\left|\sqrt{\frac{K_n}{n}}B(X_i)^T\boldsymbol\delta\{\psi(X_i,T_i,U_i)-E[\psi(X_i,T_i,U_i)|X_i]\}\right|^{2+\gamma}\bigg|X_i\right]\\
    \le& O(K_n^{-(2+\gamma)/2})O\left(n\left(\frac{K_n}{n}\right)^{(2+\gamma)/2}\right)\\
    =& o(1)
\end{align*}
since $\gamma \ge 0$. This leads to
$$\frac{Z_n-E[Z_n|X^{(n)}]}{\sqrt{V[Z_n|X^{(n)}]}}  \stackrel{D}{\rightarrow} N(0,1)$$
from Lyapunov's Theorem. The conditional expectation of $Z_n$ respect to $X^{(n)}$ can be calculated as

\begin{align*}
E[Z_n|X^{(n)}] &= -\sqrt{\frac{K_n}{n}} \sum_{i=1}^n B(x_i)^T\boldsymbol\delta E[\psi(X_i,T_i,U_i)|X_i=x_i]\\
&= -\sqrt{nK_n} \frac{1}{n}\sum_{i=1}^n p(x_i)w(x_i,1)c(x_i,1)b_a(x_i)B(x_i)^T\boldsymbol\delta  \times\\
&\bigg\{-\Phi''\big([1-\alpha^{(T_i)}]c(X_i,T_i)b_a(X_i)[1+\Tilde{o}(1)]|X_i=x_i,T_i=1\big)c(x_i,1)\\
&+\Phi''\big([1-\alpha^{(T_i)}]c(X_i,T_i)b_a(X_i)[1+\Tilde{o}(1)]|X_i=x_i,T_i=-1\big)c(x_i,-1)\bigg\}[1+\Tilde{o}(1)]\\
&= -\sqrt{nK_n} \int_0^1 p(x)w(x,1)c(x,1)b_a(x)B(x)^T\boldsymbol\delta  \times\\
    &\bigg\{-\Phi''\big([1-\alpha^{(T_i)}]c(X_i,T_i)b_a(X_i)[1+\Tilde{o}(1)]|X_i=x,T_i=1\big)c(x,1)\\
&+\Phi''\big([1-\alpha^{(T_i)}]c(X_i,T_i)b_a(X_i)[1+\Tilde{o}(1)]|X_i=x,T_i=-1\big)c(x,-1)\bigg\} dQ(x)[1+\Tilde{o}(1)]\\
    &= o\left(\sqrt{nK_n}K_n^{-(q+2)}\right).
\end{align*}
The last step is from the proof of Lemma 6.10 of \supercite{agarwal1980asymptotic} and equation (6), for $j=-p+1,...,K_n$, we have
$$\begin{aligned}
&\int_0^1 p(x)w(x,1)c(x,1)c(x,t)b_a(x)B_j(x)^T\boldsymbol\delta \times\\ &\Phi''([1-\alpha^{(T)}]c(X,T)b_a(X)[1+\Tilde{o}(1)]|X=x,T=t)dQ(x)[1+\Tilde{o}(1)]\\=&o(K_n^{-(q+2)}),
\end{aligned}$$
by which $\sqrt{nK_n}o(K_n^{-(q+2)})=o(1)$ from the order of $K_n$ in Theorem 1. Consequently, we have $E[Z_n|X^{(n)}]/\sqrt{V[Z_n|X^{(n)}]}=o_P(1)$ and Lemma 1 holds. $\blacksquare$

\bigskip
\bigskip

\noindent\textbf{Lemma 2.} Let $\nu$ be a continuous function on the interval $[0,1]$, then $\mathbf{D}=O(K_n^{-1})$. Furthermore, $\mathbf{D}^{-1}=O(K_n)$.

\noindent\textit{Proof of Lemma 2.} The $(i,j)$-component of $\mathbf{D}$ is
$$d_{ij}=\int_0^1 \nu(x)B_i(x)B_j(x)dQ(x).$$
From the fundamental property of B-spline function (Lemma 6.1 in \supercite{zhou1998local}), we have
$$|g_{ij}(\nu)|\le sup_{x\in [0,1]} \{|\nu(x)|\} sup_{x\in [0,1]} \{|Q(x)|\} max_{i,j}\int_0^1 B_i(x)B_j(x)dx=O(K_n^{-1}).$$
From the property of B-spline function \supercite{de1978practical}, $\mathbf{D}$ is positive definite matrix. Therefore, $\mathbf{D}^{-1}=O(K_n)$ is satisfied. $\blacksquare$

Now, we are ready to prove Theorem 1. For simplicity we write $a_n \stackrel{as}{\sim} b_n$, where random sequence $\{a_n\}$ and $\{b_n\}$, if $a_n/b_n=O_P(1)$. 

\noindent\textit{Proof of Theorem 1.} The objective function of proposed method is
$$\begin{aligned}
L_n(\boldsymbol\beta) &= \sum_{i=1}^n w(X_i,T_i) \rho\left(Y_i-g(X_i)-c(X_i,T_i)B(X_i)^T\boldsymbol\beta\right)\\
&= \sum_{i=1}^n w(X_i,T_i) \rho\left(Y_i-g(X_i)-c(X_i,T_i)B(X_i)^T[\boldsymbol\beta-\boldsymbol\beta^*+\boldsymbol\beta^*]\right)\\
&= \sum_{i=1}^n w(X_i,T_i) \rho\left(Y_i-g(X_i)-c(X_i,T_i)B(X_i)^T\boldsymbol\beta^*-c(X_i,T_i)B(X_i)^T[\boldsymbol\beta-\boldsymbol\beta^*]\right).\\
\end{aligned}$$

As this minimization problem doesn't have explicit solution, for the convergence of $\sqrt{a_n}(\hat{\boldsymbol\beta}-\boldsymbol\beta^*)$, we modify the objective function $L_n(\boldsymbol\beta)$ as follows:
$$
U_n(\delta) =  \sum_{i=1}^n  \left[w(X_i,T_i)\left(\rho\left(U_i-\sqrt{\frac{K_n}{n}} c(X_i,T_i) B(X_i)^T\boldsymbol\delta\right)-\rho\left(U_i\right)\right)\right],
$$
where $U_i=Y_i-g(X_i)-c(X_i,T_i)B(X_i)^T\boldsymbol\beta^*$. Then the minimizer $\hat{\delta}_n$ of $U_n(\delta)$ can be obtained as $$\hat{\delta}_n=\sqrt{\frac{n}{K_n}} (\hat{\boldsymbol\beta}-\boldsymbol\beta^*).$$ 
Define
$$R_n(\boldsymbol\delta)=U_n(\boldsymbol\delta)-E[U_n(\boldsymbol\delta)|X^{(n)},T_n]-\sum_{i=1}^n w(X_i,T_i) \{\rho'(U_i)-E[\rho'(U_i)|X_i,T_i]\}\{\alpha_n c(X_i,T_i)B(X_i)^T\boldsymbol\delta\},$$
where $X^{(n)}$ represents all the observed $X$. We have $E[R_n(\boldsymbol\delta)|X^{(n)},T_n]=0$ from the straight calculation. Let
$$r_i=w(X_i,T_i)\left[\rho\left(U_i-\alpha_n c(X_i,T_i)B(X_i)^T\boldsymbol\delta\right)-\rho(U_i)-\rho'(U_i)\{\alpha_n c(X_i,T_i)B(X_i)^T\boldsymbol\delta\}\right].$$
Then by Condition C9-10 with $s=\alpha_n c(X_i,T_i)B(X_i)^T\boldsymbol\delta$, the variance of $r_i$ is
$$\begin{aligned}
V[r_i]=&E\left[(w(X_i,T_i)\left\{\rho\left(U_i-\alpha_n c(X_i,T_i)B(X_i)^T\boldsymbol\delta\right)-\rho(U_i)-\rho'(U_i)\{\alpha_n c(X_i,T_i)B(X_i)^T\boldsymbol\delta\}\right\}\right]^2\\
&- \left[w(X_i,T_i)\{\Phi(\alpha_n c(X_i,T_i)B(X_i)^T\boldsymbol\delta|X_i,T_i)-\Phi(0|X_i,T_i)-\Phi'(0|X_i,T_i)\alpha_n c(X_i,T_i)B(X_i)^T\boldsymbol\delta\}\right]^2\\
=&o(\alpha_n^2).
\end{aligned}$$
Therefore, we have from $K_n=o(n^{1/2})$, $E[R_n(\boldsymbol\delta)^2]=\frac{1}{n}V[r_1]=o(1)$ and $R_n(\boldsymbol\delta)=o_P(1)$. By the definition of $\Phi(t|X,T)$, the Taylor expansion of $$\Phi(\alpha_n c(X_i,T_i)B(X_i)^T\boldsymbol\delta|X_i,T_i)$$ around $\alpha_n=0$, we have $E[\rho(U_i-\alpha_n c(X_i,T_i)B(X_i)^T\boldsymbol\delta)|X_i,T_i]=\Phi(\alpha_n c(X_i,T_i)B(X_i)^T\boldsymbol\delta|X_i,T_i)$ and
$$\begin{aligned}
&\Phi(\alpha_n c(X_i,T_i)B(X_i)^T\boldsymbol\delta|X_i,T_i)\\
=& \Phi(0|X_i,T_i)+\Phi'(0|X_i,T_i)\alpha_n c(X_i,T_i)B(X_i)^T\boldsymbol\delta+\frac{1}{2}\Phi''(0|X_i,T_i)\{\alpha_n c(X_i,T_i)B(X_i)^T\boldsymbol\delta\}^2+o(\alpha_n^2)
\end{aligned}$$
Therefore, the conditional expectation of $U_n(\boldsymbol\delta)$ given $X_n$ can be written as
$$\begin{aligned}
&E[U_n(\boldsymbol\delta)|X^{(n)},T_n]\\
=& \sum_{i=1}^n w(X_i,T_i)\bigg[\Phi'(0|X_i,T_i)\alpha_n c(X_i,T_i)B(X_i)^T\boldsymbol\delta+\frac{1}{2}\sum_{i=1}^n\Phi''(0|X_i,T_i)\{\alpha_n c(X_i,T_i)B(X_i)^T\boldsymbol\delta\}^2 \bigg]+o(\alpha_n^2).
\end{aligned}$$
Thus, we have $U_n(\boldsymbol\delta)$ as
$$\begin{aligned}
U_n(\boldsymbol\delta)&=E[U_n(\boldsymbol\delta|X^{(n)},T_n)]+\sum_{i=1}^n w(X_i,T_i) \{\rho'(U_i)-E[\rho'(U_i)|X_i,T_i]\}\{\alpha_n c(X_i,T_i)B(X_i)^T\boldsymbol\delta\}+o_P(1)\\
&=-\sqrt{K_n}\mathbf{W}_n^T\boldsymbol\delta+\frac{K_n}{2}\boldsymbol\delta^T \mathbf{G}_n \boldsymbol\delta+o_P(1),
\end{aligned}$$
where
$$\begin{aligned}
\mathbf{W}_n&= -\sqrt{\frac{1}{n}} \sum_{i=1}^n \rho'(U_i)w(X_i,T_i)c(X_i,T_i)B(X_i) \\
\mathbf{G}_n&= \frac{1}{n}\sum_{i=1}^n \Phi''(0|X_i,T_i)w(X_i,T_i)c(X_i,T_i)^2B(X_i)^TB(X_i).
\end{aligned}$$
The minimizer of $U_n(\boldsymbol\delta)$ is 
$$\hat{\boldsymbol\delta}=argmin_{\boldsymbol\delta}\{U_{n}(\boldsymbol\delta)\}=\mathbf{G}_n^{-1}\frac{\mathbf{W}_n}{\sqrt{K_n}}+o_P(1),$$
which is the solution of $\partial Q_{n}(\boldsymbol\delta)/\partial \boldsymbol\delta=0$. Hence, because $\hat{\boldsymbol\delta}=\frac{1}{\alpha_n} (\hat{\boldsymbol\beta}-\boldsymbol\beta^*)$, we have 
$$\begin{aligned}
\sqrt{\frac{n}{K_n}}(\hat{\tau}(x)-\tau^*(x))
=\sqrt{\frac{n}{K_n}}B(x)^T \mathbf{G}_n^{-1}\frac{\mathbf{W}_n}{\sqrt{K_n}}+o_P(1).
\end{aligned}$$
The asymptotic variance of $\hat{\tau}(x)$ is similar to that of $\hat{\tau}(x)-\tau^*(x)$ because $\mathbf{W}_n$ is the only random vector in the asymptotic form of $\hat{\tau}(x)$, it is easy to show that
$$V[\hat{\tau}(x)]=\frac{1}{K_n}B(x)^T\mathbf{G}_nV[\mathbf{W}_n]\mathbf{G}_nB(x)(1+o(1)),$$
where $\mathbf{G}_n=\mathbf{D}+o(K_n^{-1})$ and $\nu(x)=p(x)w(x,1)c(x,1)^2\rho''(y^{(1)}-g(x)-c(x,1)B(x)^T\boldsymbol\beta^*)+(1-p(x))w(x,-1)c(x,-1)^2\rho''(y^{(-1)}-g(x)-c(x,-1)B(x)^T\boldsymbol\beta^*)$ due to the Riemann integral, fundamental asymptotic property of B-spline basis, and Lemma 2. Under the condition $K_n=O(n^{1/(2q+3)})$, we have
$$\sqrt{\frac{n}{K_n}}\{\hat{\tau}(x)-\tau_0(x)\}=\sqrt{\frac{n}{K_n}}\{\hat{\tau}(x)-\tau^*(x)+b^a(x)+o(K_n^{-(q+1)})\}$$
and $\sqrt{\frac{n}{K_n}}b^a(x)=O\left(\sqrt{\frac{n}{K_n}}K_n^{-(q+1)}\right)=O(1)$. 
Thus, we have
$$\sqrt{\frac{n}{K_n}}(\hat{\tau}(x)-\tau_0(x)-b^a(x)) \stackrel{D}{\rightarrow} N(0,\Psi(x)),$$
where
$\Psi(x)=lim_{n \rightarrow \infty} \frac{1}{K_n}B(x)^T\mathbf{D}^{-1}\mathbf{G}\mathbf{D}^{-1}B(x).$ This completes the proof. $\blacksquare$

To prove Theorem 2, we introduce two lemmas as well.

\noindent\textbf{Lemma 3.} Let $U_i=w(X_i,T_i)(Y_i-g(X_i)-c(X_i,T_i)B(X_i)^T\boldsymbol\beta^*)$. Under the same assumptions as Theorem 2,
$$-\sqrt{\frac{K_n}{n}}\sum_{i=1}^n w(X_i,T_i)c(X_i,T_i)B(X_i)^T \boldsymbol\delta [1-2I(U_i<0)] \stackrel{as}{\sim} -\sqrt{K_n} \mathbf{W}^T \boldsymbol\delta,$$
where $\mathbf{W} \sim N(0,\mathbf{G})$.

\noindent\textit{Proof of Lemma 3.} 
Let $Z_n=-\sqrt{\frac{K_n}{n}}\sum_{i=1}^n w(X_i,T_i)c(X_i,T_i)B(X_i)^T\boldsymbol\delta[1-2I(U_i<0)]$,
the conditional expectation of $w(X_i,T_i)c(X_i,T_i)[1-2I(U_i<0)]$ respect to $X_i$ can be calculated as following.
\begin{align*}
&E[w(X_i,T_i)c(X_i,T_i)[1-2I(U_i<0)]|X_i=x_i] \\
=& p(x_i)w(x_i,1)c(x_i,1)E[1-2I(U_i<0)|X_i=x_i,T_i=1]\\
&+(1-p(x_i))w(x_i,-1)c(x_i,-1)E[1-2I(U_i<0)|X_i=x_i,T_i=-1]\\
=& p(x_i)w(x_i,1)c(x_i,1)\{1-2E[I(U_i<0)|X_i=x_i,T_i=1]-1+2E[I(U_i<0)|X_i=x_i,T_i=-1]\}\\
=& 2p(x_i)w(x_i,1)c(x_i,1)[P(U_i<0|X_i=x_i,T_i=-1)-P(U_i<0|X_i=x_i,T_i=1)]\\
=& 2p(x_i)w(x_i,1)c(x_i,1)[P(Y_i^{(-1)}<g(x_i)+c(x_i,-1)B(x_i)^T\beta^*)-P(Y_i^{(1)}<g(x_i)+c(x_i,1)B(x_i)^T\boldsymbol\beta^*)]\\
=& 2p(x_i)w(x_i,1)c(x_i,1)[P(\varepsilon_i^{(-1)}<g(x_i)-b(x_i)+[c(x_i,1)-0.5]\tau_0(x_i)+[c(x_i,1)-1]b^a(x_i)(1+o(1))|x_i)\\
&-P(\varepsilon_i^{(1)}<g(x_i)-b(x_i)+[c(x_i,1)-0.5]\tau_0(x_i)+c(x_i,1)b^a(x_i)(1+o(1))|x_i)]\\
=& -2p(x_i)w(x_i,1)c(x_i,1)b^a(x_i)f_{\varepsilon_i}(g(x_i)-b(x_i)+[c(x_i,1)-0.5]\tau_0(x_i)|x_i)(1+o(1))\\
=& o(1).
\end{align*}
The derivation from the first equation to the second equation is by Condition C1, that from the third equation to the fourth equation is by Model (1), that to the fifth equation is by proposed Condition C2, the last two steps are by Taylor expansion and the order of $b^a(x_i)$.

Therefore, let $\psi(X_i,T_i,U_i)=w(X_i,T_i)c(X_i,T_i)[1-2I(U_i<0)]$, we obtain
$$\begin{aligned}
&E\left[\left|\sqrt{\frac{K_n}{n}}B(X_i)^T\boldsymbol\delta \left[\psi(X_i,T_i,U_i)-E[\psi(X_i,T_i,U_i)|X_i=x_i]\right]\right|^{2+\gamma}\bigg|X_i=x_i\right]\\
=& \left(\frac{K_n}{n}\right)^{\frac{2+\gamma}{2}} |B(x_i)^T\boldsymbol\delta|^{2+\gamma} E\left[\left|\psi(X_i,T_i,U_i)\right|^{2+\gamma}+o(1)\bigg|X_i=x_i\right]\\
=& \left(\frac{K_n}{n}\right)^{\frac{2+\gamma}{2}} |B(x_i)^T\boldsymbol\delta|^{2+\gamma} \{p(x_i)E[|w(x_i,1)c(x_i,1)[1-2I(U_i<0)]|^{2+\gamma}+o(1)|X_i=x_i,T_i=1]\\
&+(1-p(x_i))E[|w(x_i,-1)c(x_i,-1)[1-2I(U_i<0)]|^{2+\gamma}+o(1)|X_i=x_i,T_i=-1]\}\\
\le& O\left(\left(\frac{K_n}{n}\right)^{\frac{2+\gamma}{2}}\right),
\end{aligned}$$

where the last two steps are derived by Condition C12. The conditional variance of $Z_n$ respect to $X^{(n)}$ can be calculated as following.
$$\begin{aligned}
V[Z_n|X^{(n)}]&= \frac{K_n}{n} \sum_{i=1}^n \{B(x_i)^T\boldsymbol\delta\}^2 V\left[w(X_i,T_i)c(X_i,T_i)[1-2I(U_i<0)]\bigg|X_i=x_i\right]\\
&= \frac{K_n}{n} \sum_{i=1}^n \{B(x_i)^T\boldsymbol\delta\}^2 \bigg\{E\left[\left(w(X_i,T_i)c(X_i,T_i)[1-2I(U_i<0)]\right)^2\bigg|X_i=x_i\right] \\
&- E\left[w(X_i,T_i)c(X_i,T_i)[1-2I(U_i<0)]\bigg|X_i=x_i \right]^2 \bigg\}\\
&= \frac{K_n}{n} \sum_{i=1}^n \{B(x_i)^T\boldsymbol\delta\}^2 \bigg\{p(x_i)w(x_i,1)^2c(x_i,1)^2+(1-p(x_i))w(x_i,-1)^2c(x_i,-1)^2 \\
&- E\left[w(X_i,T_i)c(X_i,T_i)[1-2I(U_i<0)]\bigg|X_i=x_i \right]^2 \bigg\}\\
&= \frac{K_n}{n} \sum_{i=1}^n \{B(x_i)^T\boldsymbol\delta\}^2 \bigg\{\frac{p(x_i)}{(1-p(x_i))}w(x_i,1)^2c(x_i,1)^2\bigg\}(1+o_P(1))\\
&= K_n\boldsymbol\delta^T\mathbf{G}\boldsymbol\delta(1+o_P(1))\\
&= O(K_n).
\end{aligned}$$

So it follows that
\begin{align*}
    \frac{1}{V[Z_n|X^{(n)}]^{(2+\gamma)/2}}\sum_{i=1}^n &E\left[\left|\sqrt{\frac{K_n}{n}}B(X_i)^T\boldsymbol\delta\{\psi(X_i,T_i,U_i)-E[\psi(X_i,T_i,U_i)|X_i]\}\right|^{2+\gamma}\bigg|X_i\right]\\
    \le& O(K_n^{-(2+\gamma)/2})O\left(n\left(\frac{K_n}{n}\right)^{(2+\gamma)/2}\right)\\
    =& o(1)
\end{align*}
since $\gamma \ge 0$. This leads to
$$\frac{Z_n-E[Z_n|X^{(n)}]}{\sqrt{V[Z_n|X^{(n)}]}}  \stackrel{D}{\rightarrow} N(0,1)$$
from Lyapunov's Theorem. The conditional expectation of $Z_n$ respect to $X^{(n)}$ can be calculated as

\begin{align*}
    E[Z_n|X^{(n)}] &= -\sqrt{\frac{K_n}{n}} \sum_{i=1}^n B(X_i)^T\boldsymbol\delta E[\psi(X_i,T_i,U_i)|X_i]\\
    &= 2\sqrt{\frac{K_n}{n}} \sum_{i=1}^n p(x_i)w(x_i,1)c(x_i,1)b^a(x_i)f_{\varepsilon_i}(g(x_i)-b(x_i)+[c(x_i,1)-0.5]\tau_0(x_i))(1+o(1))\\
    &= 2\sqrt{nK_n} \int_0^1 p(u)w(u,1)c(u,1)b^a(u)f_{\varepsilon_i}(g(u)-b(u)+[c(u,1)-0.5]\tau_0(u)|u)dQ(u)(1+o(1)).
\end{align*}
From the proof of Lemma 6.10 of \supercite{agarwal1980asymptotic}, for $j=-p+1,...,K_n$, we have
$$\int_0^1 p(u)w(u,1)c(u,1)b^a(u)f_{\varepsilon_i}(g(u)-b(u)+[c(u,1)-0.5]\tau_0(u)|u)dQ(u)(1+o(1))=o(K_n^{-(p+2)}),$$
by which $\sqrt{nK_n}o(K_n^{-(p+2)})=o(1)$. Consequently, we have $E[Z_n|X^{(n)}]/\sqrt{V[Z_n|X^{(n)}]}=o_P(1)$ and Lemma 3 holds. $\blacksquare$

\noindent\textbf{Lemma 4.} Let $w_{in}=\sqrt{\frac{K_n}{n}} w(x_i,t_i)c(x_i,t_i) B(x_i)^T \boldsymbol\delta (i=1,...,n)$ for $\boldsymbol\delta \in \mathbb{R}^{K_n+q}$. Then, under the assumptions of Theorem 2,
$$2\sum_{i=1}^n \int_0^{w_{in}} [I(U_i \le s)-I(U_i \le 0)]ds \stackrel{as}{\sim} K_n\boldsymbol\delta^T \mathbf{D} \boldsymbol\delta.$$

\noindent\textit{Proof of Lemma 4.} Let
$$R_n=2\sum_{i=1}^n \int_0^{w_{in}} [I(U_i \le s)-I(U_i \le 0)]ds.$$
Since 
$$\begin{aligned}
&E[\int_0^{w_{in}} [I(U_i\le s)-I(U_i \le 0)]ds|X_i=x_i,T_i=t_i]\\
=& \int_0^{w_{in}} E[I(U_i\le s)-I(U_i \le 0)|X_i=x_i,T_i=t_i] ds\\
=& \int_0^{w_{in}} E\left[I\left(Y_i\le g(X_i)+c(X_i,T_i)\tau^*(X_i)+\frac{s}{w(X_i,T_i)}\right)-I\left(Y_i\le g(X_i)+c(X_i,T_i)\tau^*(X_i)\right)\bigg|X_i=x_i,T_i=t_i\right]ds\\
=& \int_0^{w_{in}} P\left(Y_i\le g(X_i)+c(X_i,T_i)\tau^*(X_i)+\frac{s}{w(X_i,T_i)}\bigg|X_i=x_i,T_i=t_i\right)\\
&-P\left(Y_i\le g(X_i)+c(X_i,T_i)\tau^*(X_i)\bigg|X_i=x_i,T_i=t_i\right) ds\\
=& \sqrt{\frac{K_n}{n}}w(x_i,t_i)c(x_i,t_i) \int_0^{B(x_i)^T\boldsymbol\delta} P\left(Y_i\le g(X_i)+c(X_i,T_i)\tau^*(X_i)+\sqrt{\frac{K_n}{n}}c(X_i,T_i)t\bigg|X_i=x_i,T_i=t_i\right)\\
&-P\left(Y_i\le g(X_i)+c(X_i,T_i)\tau^*(X_i)\bigg|X_i=x_i,T_i=t_i\right) dt\\
=& \frac{K_n}{n} w(x_i,t_i)c(x_i,t_i)^2 \int_0^{B(x_i)^T\boldsymbol\delta} f\left(g(X_i)+c(X_i,T_i)\tau^*(X_i)\bigg|X_i=x_i,T_i=t_i\right)tdt (1+o(1))\\
=&  \frac{K_n}{2n} w(x_i,t_i)c(x_i,t_i)^2 f\left(g(X_i)+c(X_i,T_i)\tau^*(X_i)\bigg|X_i=x_i,T_i=t_i\right)\{B(x_i)^T\boldsymbol\delta\}^2(1+o(1)).
\end{aligned}$$

Therefore we obtain
$$\begin{aligned}
E[R_n|X^{(n)}]&= 2 \frac{K_n}{2n} \sum_{i=1}^n \bigg\{p(x_i)w(x_i,1)c(x_i,1)^2   f_1\left(g(X_i)+c(X_i,T_i)\tau^*(X_i)\bigg|X_i=x_i,T_i=1\right)\\
+& (1-p(x_i))w(x_i,-1)c(x_i,-1)^2f_{-1}\left(g(X_i)+c(X_i,T_i)\tau^*(X_i)\bigg|X_i=x_i,T_i=-1\right)\bigg\} \boldsymbol\delta^TB(x_i)B(x_i)^T\boldsymbol\delta(1+o(1))\\
&= K_n \boldsymbol\delta^T \bigg\{\frac{1}{n}\sum_{i=1}^n  [p(x_i)w(x_i,1)c(x_i,1)^2f_1(g(x_i)+c(x_i,1)\tau_0(x_i)|x_i)\\
+& (1-p(x_i))w(x_i,-1)c(x_i,-1)^2 f_{-1}(g(x_i)+c(x_i,-1)\tau_0(x_i)|x_i)]B(x_i)B(x_i)^T\bigg\}\boldsymbol\delta (1+o_P(1))\\
&= K_n \boldsymbol\delta^T \mathbf{D} \boldsymbol\delta(1+o_P(1)).
\end{aligned}$$

For $i=1,...,n$, we have
$$\int_0^{w_{in}} [I(u_i\le s)-I(u_i \le 0)]ds \le \sqrt{\frac{K_n}{n}} w(x_i,t_i)c(x_i,t_i) B(x_i)^T\boldsymbol\delta.$$
Therefore the variance of $R_n$ can be evaluated as
$$\begin{aligned}
V[R_n|X^{(n)}] &\le \sum_{i=1}^n E\left[\left(\int_0^{w_{in}} [I(u_i\le s)-I(u_i \le 0)]ds\right)^2\bigg|X_i=x_i\right]\\
&\le \sqrt{\frac{K_n}{n}} max_{i=1,...,n}\left\{w(x_i,t_i)c(x_i,t_i)\right\} E[R_n|X^{(n)}].
\end{aligned}$$
Since $E[R_n|X^{(n)}]=O(K_n)$, we obtain $\sqrt{V[R_n|X^{(n)}]}/E[R_n|X^{(n)}]=o_P(1)$ and hence, Lemma 4 holds.$\blacksquare$

Now we are ready to prove Theorem~2. 

\noindent\textit{Proof of Theorem 2.} The objective function of proposed method is
$$\begin{aligned}
L(\boldsymbol\beta) &= \sum_{i=1}^n w(X_i,T_i) \left|Y_i-g(X_i)-c(X_i,T_i)B(X_i)^T\boldsymbol\beta\right|\\
&= \sum_{i=1}^n w(X_i,T_i) \left|Y_i-g(X_i)-c(X_i,T_i)B(X_i)^T(\boldsymbol\beta-\boldsymbol\beta^*+\boldsymbol\beta^*)\right|\\
&= \sum_{i=1}^n w(X_i,T_i) \left|Y_i-g(X_i)-c(X_i,T_i)B(X_i)^T\boldsymbol\beta^*-c(X_i,T_i)B(X_i)^T(\boldsymbol\beta-\boldsymbol\beta^*)\right|.\\
\end{aligned}$$

Let
$$
U_n(\boldsymbol\delta) =  \sum_{i=1}^n  \left[\left|U_i-\sqrt{\frac{K_n}{n}} w(X_i,T_i)c(X_i,T_i) B(X_i)^T\boldsymbol\delta\right|-\bigg|U_i\bigg|\right]=U_{1n}(\boldsymbol\delta)+U_{2n}(\boldsymbol\delta),
$$
where $U_i=w(X_i,T_i)\left(Y_i-g(X_i)-c(X_i,T_i)B(X_i)^T\boldsymbol\beta^*\right)$. Then the minimizer $\hat{\boldsymbol\delta}_n$ of $U_n(\boldsymbol\delta)$ can be obtained as $$\hat{\boldsymbol\delta}_n=\sqrt{\frac{n}{K_n}} (\hat{\boldsymbol\beta}-\boldsymbol\beta^*).$$ Following the Knight's identity, we can write $U_n(\boldsymbol\delta)$ as
$$U_n(\boldsymbol\delta)=U_{1n}(\boldsymbol\delta)+U_{2n}(\boldsymbol\delta),$$
where
$$\begin{aligned}
U_{1n}(\boldsymbol\delta)&= -\sqrt{\frac{K_n}{n}} \sum_{i=1}^n w(X_i,T_i)c(X_i,T_i)B(X_i)^T\boldsymbol\delta [1-2I(U_i<0)]\\
U_{2n}(\boldsymbol\delta)&= 2\sum_{i=1}^n \int_0^{w_{in}} I(U_i\le s)-I(U_i\le 0)ds,
\end{aligned}$$
where $w_{in}=\sqrt{\frac{K_n}{n}} w(x_i,t_i)c(x_i,t_i) B(x_i)^T\boldsymbol\delta$. From Lemma 3,
$$U_{1n}(\boldsymbol\delta) \stackrel{as}{\sim} -\sqrt{K_n}\mathbf{W}^T\boldsymbol\delta,$$
where $\mathbf{W} \sim N(0,\mathbf{G})$. Furthermore, Lemma 4 yield
$$U_{2n}(\boldsymbol\delta) \stackrel{as}{\sim} K_n \boldsymbol\delta^T\mathbf{D}\boldsymbol\delta.$$

Therefore, for both methods we obtain
$$U_n(\boldsymbol\delta) \stackrel{as}{\sim} U_{0n}(\boldsymbol\delta)=-\sqrt{K_n}\mathbf{W}^T \boldsymbol\delta+K_n \boldsymbol\delta^T \mathbf{D} \boldsymbol\delta.$$
Because $U_{0n}(\boldsymbol\delta)$ is convex with respect to $\boldsymbol\delta$ and has unique minimizer, the minimizer of $U_{n}(\boldsymbol\delta)$ converges to $\boldsymbol\delta_0=argmin_{\boldsymbol\delta}\{U_{0n}(\boldsymbol\delta)\}$. This fact is detailed in \supercite{knight1998limiting}. Hence, we have
$$\sqrt{\frac{n}{K_n}} \{\hat{\boldsymbol\beta}-\boldsymbol\beta^*\} \stackrel{as}{\sim} \boldsymbol\delta_0 = \mathbf{D}^{-1}\left(\frac{1}{2\sqrt{K_n}}\mathbf{W}\right).$$
Since $\hat{\tau}(x)-\tau^*(x)=B(x)^T (\hat{\boldsymbol\beta}-\boldsymbol\beta^*)$, we obtain for $x \in (0,1)$, as $n \rightarrow \infty$,
$$\sqrt{\frac{n}{K_n}}\{\hat{\tau}(x)-\tau^*(x)\} \stackrel{D}{\rightarrow} N(0,\Psi(x)),$$
where $\Psi(x)=lim_{n \rightarrow \infty} \frac{1}{4K_n}B(x)^T\mathbf{D}^{-1}\mathbf{G}\mathbf{D}^{-1}B(x)$ by the definition of $\mathbf{W}$. 

Under the condition $K_n=O(n^{1/(2q+3)})$, we have
$$\sqrt{\frac{n}{K_n}}\{\hat{\tau}(x)-\tau_0(x)\}=\sqrt{\frac{n}{K_n}}\{\hat{\tau}(x)-\tau^*(x)-b^a(x)+o(K_n^{-(q+1)})\}$$
and $\sqrt{\frac{n}{K_n}}b^a(x)=O\left(\sqrt{\frac{n}{K_n}}K_n^{-(q+1)}\right)=O(1)$. This completes the proof. $\blacksquare$

\newpage
\subsection*{B. Simulation Results}
\subsubsection*{B.1. Simulation Results of Settings 1-3}

\begin{table}[H]
\begin{threeparttable}
\caption*{Table B.1: Simulation Results for Setting 1. In the presence of outliers, $L_1$-MCMEA and $L_1$-RL outperformed their $L_2$-based counterparts. The MSE and MAE decreased and sensitivity, specificity, and $Q(\hat{\eta})$ increased with sample size.}
\begin{tabular}{ |l|l|c|c|c|c|c|c|c| } 
\hline
$\xi_o$ &Measurement	&	Bias.sq	&	Var	&	MSE	&	MAE	&	Sensitivity	&	Specificity	&	$Q(\hat{\eta})$	\\
\hline \hline
\multirow{8}{*}{0}&	MCMEA	&	0.34	&	0.08	&	0.42	&	0.48	&	1.00	&	0.49	&	1.18	\\
&	$L_1$-MCMEA	&	0.27	&	0.23	&	0.50	&	0.52	&	1.00	&	0.88	&	1.18	\\
&	RL	&	0.18	&	0.24	&	0.42	&	0.47	&	1.00	&	0.60	&	1.20	\\
&	$L_1$-RL	&	0.14	&	0.32	&	0.46	&	0.50	&	1.00	&	0.93	&	1.19	\\
&	AL	&	0.35	&	0.19	&	0.54	&	0.47	&	1.00	&	0.53	&	1.18	\\
&	$L_1$-AL	&	0.28	&	0.37	&	0.64	&	0.52	&	1.00	&	0.86	&	1.18	\\
&	QL	&	0.35	&	0.16	&	0.51	&	0.43	&	1.00	&	0.30	&	1.17	\\
&	$L_1$-QL	&	0.16	&	0.28	&	0.44	&	0.46	&	1.00	&	0.41	&	1.19	\\
\hline
\multirow{8}{*}{0.05}&	MCMEA	&	0.92	&	0.59	&	1.51	&	0.95	&	0.90	&	0.50	&	1.04	\\
&	$L_1$-MCMEA	&	0.27	&	0.33	&	0.61	&	0.53	&	1.00	&	0.86	&	1.17	\\
&	RL	&	0.55	&	0.79	&	1.34	&	0.88	&	0.94	&	0.49	&	1.07	\\
&	$L_1$-RL	&	0.15	&	0.41	&	0.55	&	0.55	&	1.00	&	0.91	&	1.18	\\
&	AL	&	0.91	&	0.69	&	1.60	&	0.97	&	0.88	&	0.54	&	1.03	\\
&	$L_1$-AL	&	0.32	&	0.47	&	0.79	&	0.55	&	1.00	&	0.85	&	1.17	\\
&	QL	&	0.76	&	0.54	&	1.30	&	0.88	&	0.96	&	0.46	&	1.07	\\
&	$L_1$-QL	&	0.17	&	0.31	&	0.48	&	0.48	&	1.00	&	0.43	&	1.18	\\
\hline
\multirow{8}{*}{0.1}&	MCMEA	&	1.41	&	0.75	&	2.16	&	1.17	&	0.72	&	0.62	&	0.91	\\
&	$L_1$-MCMEA	&	0.28	&	0.42	&	0.70	&	0.56	&	1.00	&	0.85	&	1.16	\\
&	RL	&	1.08	&	0.99	&	2.08	&	1.13	&	0.77	&	0.59	&	0.94	\\
&	$L_1$-RL	&	0.15	&	0.45	&	0.61	&	0.57	&	1.00	&	0.92	&	1.18	\\
&	AL	&	1.39	&	0.88	&	2.27	&	1.19	&	0.70	&	0.64	&	0.90	\\
&	$L_1$-AL	&	0.34	&	0.60	&	0.94	&	0.60	&	0.99	&	0.81	&	1.16	\\
&	QL	&	1.20	&	0.66	&	1.86	&	1.08	&	0.86	&	0.54	&	0.98	\\
&	$L_1$-QL	&	0.17	&	0.33	&	0.50	&	0.50	&	1.00	&	0.48	&	1.18	\\
\hline
\multirow{8}{*}{0.15}&	MCMEA	&	1.72	&	0.88	&	2.60	&	1.29	&	0.58	&	0.69	&	0.81	\\
&	$L_1$-MCMEA	&	0.30	&	0.51	&	0.81	&	0.61	&	0.99	&	0.85	&	1.15	\\
&	RL	&	1.48	&	1.12	&	2.60	&	1.28	&	0.62	&	0.69	&	0.83	\\
&	$L_1$-RL	&	0.15	&	0.52	&	0.67	&	0.60	&	0.99	&	0.91	&	1.17	\\
&	AL	&	1.71	&	0.99	&	2.69	&	1.31	&	0.56	&	0.71	&	0.80	\\
&	$L_1$-AL	&	0.39	&	0.76	&	1.15	&	0.67	&	0.97	&	0.83	&	1.15	\\
&	QL	&	1.15	&	0.78	&	2.23	&	1.19	&	0.79	&	0.56	&	0.91	\\
&	$L_1$-QL	&	0.18	&	0.38	&	0.56	&	0.53	&	0.99	&	0.47	&	1.17	\\
\hline
\multirow{8}{*}{0.2}&	MCMEA	&	1.98	&	0.98	&	2.96	&	1.39	&	0.46	&	0.76	&	0.73	\\
&	$L_1$-MCMEA	&	0.36	&	0.67	&	1.03	&	0.67	&	0.98	&	0.83	&	1.14	\\
&	RL	&	1.79	&	1.20	&	2.99	&	1.39	&	0.49	&	0.75	&	0.75	\\
&	$L_1$-RL	&	0.16	&	0.59	&	0.75	&	0.63	&	0.99	&	0.89	&	1.17	\\
&	AL	&	1.95	&	1.10	&	3.06	&	1.40	&	0.44	&	0.77	&	0.73	\\
&	$L_1$-AL	&	0.50	&	1.00	&	1.50	&	0.75	&	0.96	&	0.81	&	1.13	\\
&	QL	&	1.70	&	0.88	&	2.59	&	1.29	&	0.70	&	0.61	&	0.84	\\
&	$L_1$-QL	&	0.19	&	0.42	&	0.61	&	0.54	&	0.99	&	0.52	&	1.17	\\
\hline
\end{tabular}

  \end{threeparttable}
\end{table}

\begin{table}[H]
\begin{threeparttable}
\caption*{Table B.2: Simulation results of Setting 2. In the presence of outliers, $L_1$-MCMEA and $L_1$-RL outperformed their $L_2$-based counterparts. The MSE and MAE decreased and sensitivity and specificity increased with sample size. }
\begin{tabular}{ |l|l|l|c|c|c|c|c|c|c| } 
\hline
$\xi_o$ & Sample Size	&Method		&	Bias.sq	&	Var	&	MSE	&	MAE	&	Sensitivity	&	Specificity & $Q(\hat{\eta})$	\\
\hline\hline
\multirow{24}{*}{0.05}& \multirow{8}{*}{200}	&	MCMEA	&	2.29	&	0.92	&	3.21	&	1.43	&	0.32	&	0.83	&	0.66	\\
&	&	$L_1$-MCMEA	&	1.32	&	1.34	&	2.65	&	1.20	&	0.72	&	0.62	&	0.92	\\
&	&	RL	&	1.76	&	1.85	&	3.61	&	1.47	&	0.39	&	0.82	&	0.71	\\
&	&	$L_1$-RL	&	0.72	&	2.58	&	3.30	&	1.34	&	0.84	&	0.52	&	0.97	\\
&	&	AL	&	2.28	&	1.49	&	3.77	&	1.52	&	0.26	&	0.88	&	0.64	\\
&	&	$L_1$-AL	&	1.31	&	2.77	&	4.08	&	1.40	&	0.60	&	0.67	&	0.86	\\
&	&	QL	&	2.14	&	0.79	&	2.92	&	1.37	&	0.69	&	0.57	&	0.78	\\
&	&	$L_1$-QL	&	0.64	&	1.19	&	1.83	&	0.99	&	0.89	&	0.37	&	1.02	\\
\cline{2-10}
&\multirow{8}{*}{500}	&	MCMEA	&	1.49	&	0.75	&	2.24	&	1.18	&	0.69	&	0.65	&	0.88	\\
&	&	$L_1$-MCMEA	&	0.50	&	0.65	&	1.14	&	0.74	&	0.97	&	0.71	&	1.11	\\
&	&	RL	&	0.98	&	1.17	&	2.15	&	1.14	&	0.73	&	0.64	&	0.91	\\
&	&	$L_1$-RL	&	0.26	&	0.87	&	1.14	&	0.79	&	0.98	&	0.77	&	1.14	\\
&	&	AL	&	1.48	&	0.97	&	2.44	&	1.23	&	0.63	&	0.70	&	1.10	\\
&	&	$L_1$-AL	&	0.53	&	1.04	&	0.82	&	0.82	&	0.96	&	0.67	&	0.91	\\
&	&	QL	&	1.26	&	0.72	&	1.97	&	1.11	&	0.87	&	0.50	&	0.95	\\
&	&	$L_1$-QL	&	0.25	&	0.50	&	0.75	&	0.62	&	0.99	&	0.46	&	1.15	\\
\cline{2-10}
&\multirow{8}{*}{1000}	&	MCMEA	&	0.92	&	0.59	&	1.51	&	0.95	&	0.90	&	0.50	&	1.04	\\
&	&	$L_1$-MCMEA	&	0.27	&	0.33	&	0.61	&	0.53	&	1.00	&	0.86	&	1.17	\\
&	&	RL	&	0.55	&	0.79	&	1.34	&	0.88	&	0.94	&	0.49	&	1.07	\\
&	&	$L_1$-RL	&	0.15	&	0.41	&	0.55	&	0.55	&	1.00	&	0.91	&	1.18	\\
&	&	AL	&	0.91	&	0.69	&	1.60	&	0.97	&	0.88	&	0.54	&	1.03	\\
&	&	$L_1$-AL	&	0.32	&	0.47	&	0.79	&	0.55	&	1.00	&	0.85	&	1.17	\\
&	&	QL	&	0.76	&	0.54	&	1.30	&	0.88	&	0.96	&	0.46	&	1.07	\\
&	&	$L_1$-QL	&	0.17	&	0.31	&	0.48	&	0.48	&	1.00	&	0.43	&	1.18	\\
\hline
\multirow{24}{*}{0}&\multirow{8}{*}{200}	&	MCMEA	&	1.22	&	0.73	&	1.95	&	1.04	&	0.84	&	0.61	&	1.00	\\
&	&	$L_1$-MCMEA	&	1.20	&	1.10	&	2.31	&	1.11	&	0.83	&	0.57	&	0.98	\\
&	&	RL	&	0.45	&	0.96	&	1.41	&	0.85	&	0.98	&	0.62	&	1.10	\\
&	&	$L_1$-RL	&	0.73	&	2.23	&	2.96	&	1.26	&	0.88	&	0.53	&	1.01	\\
&	&	AL	&	1.22	&	1.62	&	2.84	&	1.17	&	0.74	&	0.70	&	0.94	\\
&	&	$L_1$-AL	&	1.19	&	2.45	&	3.64	&	1.29	&	0.75	&	0.57	&	0.93	\\
&	&	QL	&	1.34	&	0.33	&	1.67	&	1.01	&	0.93	&	0.70	&	1.03	\\
&	&	$L_1$-QL	&	0.56	&	1.03	&	1.59	&	0.93	&	0.92	&	0.34	&	1.05	\\
\cline{2-10}
&\multirow{8}{*}{500}	&	MCMEA	&	0.57	&	0.26	&	0.83	&	0.65	&	1.00	&	0.51	&	1.15	\\
&	&	$L_1$-MCMEA	&	0.44	&	0.50	&	0.94	&	0.68	&	0.99	&	0.74	&	1.14	\\
&	&	RL	&	0.26	&	0.25	&	0.51	&	0.48	&	1.00	&	0.62	&	1.18	\\
&	&	$L_1$-RL	&	0.28	&	0.76	&	1.04	&	0.74	&	0.99	&	0.81	&	1.14	\\
&	&	AL	&	0.55	&	0.42	&	0.96	&	0.64	&	0.99	&	0.55	&	1.15	\\
&	&	$L_1$-AL	&	0.47	&	0.80	&	1.27	&	0.73	&	0.97	&	0.67	&	1.13	\\
&	&	QL	&	0.48	&	0.29	&	0.76	&	0.55	&	1.00	&	0.31	&	1.14	\\
&	&	$L_1$-QL	&	0.24	&	0.45	&	0.69	&	0.59	&	0.99	&	0.42	&	1.16	\\
\cline{2-10}
&\multirow{8}{*}{1000}	&	MCMEA	&	0.38	&	0.14	&	0.52	&	0.50	&	1.00	&	0.49	&	1.18	\\
&	&	$L_1$-MCMEA	&	0.27	&	0.27	&	0.54	&	0.52	&	1.00	&	0.88	&	1.18	\\
&	&	RL	&	0.18	&	0.13	&	0.31	&	0.37	&	1.00	&	0.60	&	1.20	\\
&	&	$L_1$-RL	&	0.14	&	0.36	&	0.50	&	0.52	&	1.00	&	0.93	&	1.19	\\
&	&	AL	&	0.35	&	0.19	&	0.54	&	0.47	&	1.00	&	0.53	&	1.18	\\
&	&	$L_1$-AL	&	0.28	&	0.37	&	0.64	&	0.52	&	1.00	&	0.86	&	1.18	\\
&	&	QL	&	0.35	&	0.16	&	0.51	&	0.43	&	1.00	&	0.30	&	1.17	\\
&	&	$L_1$-QL	&	0.16	&	0.28	&	0.44	&	0.46	&	1.00	&	0.41	&	1.19	\\
\hline

\end{tabular}

  \end{threeparttable}
\end{table}

\begin{table}[H]
\begin{threeparttable}
\caption*{Table B.3. Simulation results of Setting 3. Effects of the covariate dimension. With the presence of outliers,  the $L_1$-based methods outperformed the $L_2$-based methods in MSE and MAE.}
\begin{tabular}{ |l|l|l|c|c|c|c|c|c|c| } 
\hline
$\xi_o$ & Dimension	&Method		&	Bias.sq	&	Var	&	MSE	&	MAE	&	Sensitivity	&	Specificity	& $Q(\hat{\eta})$\\
\hline\hline
\multirow{24}{*}{0.05}&	\multirow{8}{*}{10}	&	MCMEA	&	0.92	&	0.59	&	1.51	&	0.95	&	0.90	&	0.50	&	1.04	\\
&	&	$L_1$-MCMEA	&	0.27	&	0.33	&	0.61	&	0.53	&	1.00	&	0.86	&	1.17	\\
&	&	RL	&	0.55	&	0.79	&	1.34	&	0.88	&	0.94	&	0.49	&	1.07	\\
&	&	$L_1$-RL	&	0.15	&	0.41	&	0.55	&	0.55	&	1.00	&	0.91	&	1.18	\\
&	&	AL	&	0.91	&	0.69	&	1.60	&	0.97	&	0.88	&	0.54	&	1.03	\\
&	&	$L_1$-AL	&	0.32	&	0.47	&	0.79	&	0.55	&	1.00	&	0.85	&	1.17	\\
&	&	QL	&	0.76	&	0.54	&	1.30	&	0.88	&	0.96	&	0.46	&	1.07	\\
&	&	$L_1$-QL	&	0.17	&	0.31	&	0.48	&	0.48	&	1.00	&	0.43	&	1.18	\\
\cline{2-10}
&	\multirow{8}{*}{30}	&	MCMEA	&	1.06	&	0.61	&	1.67	&	1.05	&	0.87	&	0.73	&	1.01	\\
&	&	$L_1$-MCMEA	&	0.31	&	0.37	&	0.68	&	0.56	&	0.99	&	0.99	&	1.17	\\
&	&	RL	&	0.62	&	0.89	&	1.51	&	0.98	&	0.85	&	0.75	&	0.99	\\
&	&	$L_1$-RL	&	0.18	&	0.42	&	0.60	&	0.59	&	1.00	&	0.98	&	1.18	\\
&	&	AL	&	1.06	&	0.72	&	1.78	&	1.12	&	0.78	&	0.76	&	0.98	\\
&	&	$L_1$-AL	&	0.33	&	0.50	&	0.83	&	0.58	&	0.99	&	0.98	&	1.16	\\
&	&	QL	&	0.87	&	0.61	&	1.48	&	0.96	&	0.87	&	0.72	&	1.00	\\
&	&	$L_1$-QL	&	0.25	&	0.33	&	0.58	&	0.50	&	0.99	&	0.99	&	1.17	\\
\cline{2-10}
&	\multirow{8}{*}{50}	&	MCMEA	&	1.14	&	0.63	&	1.77	&	1.12	&	0.80	&	0.82	&	1.01	\\
&	&	$L_1$-MCMEA	&	0.32	&	0.43	&	0.75	&	0.59	&	0.99	&	0.99	&	1.17	\\
&	&	RL	&	0.69	&	0.92	&	1.61	&	1.05	&	0.79	&	0.81	&	0.98	\\
&	&	$L_1$-RL	&	0.21	&	0.44	&	0.65	&	0.61	&	0.99	&	0.98	&	1.18	\\
&	&	AL	&	1.10	&	0.75	&	1.85	&	1.17	&	0.73	&	0.84	&	0.96	\\
&	&	$L_1$-AL	&	0.33	&	0.55	&	0.88	&	0.62	&	0.98	&	0.99	&	1.15	\\
&	&	QL	&	0.96	&	0.62	&	1.58	&	1.04	&	0.83	&	0.75	&	0.98	\\
&	&	$L_1$-QL	&	0.29	&	0.34	&	0.63	&	0.52	&	0.98	&	0.99	&	1.17	\\
\hline
\multirow{24}{*}{0}&	\multirow{8}{*}{10}		&	MCMEA	&	0.34	&	0.08	&	0.42	&	0.48	&	1.00	&	0.49	&	1.18	\\
&	&	$L_1$-MCMEA	&	0.27	&	0.23	&	0.50	&	0.52	&	1.00	&	0.88	&	1.18	\\
&	&	RL	&	0.18	&	0.24	&	0.42	&	0.47	&	1.00	&	0.60	&	1.20	\\
&	&	$L_1$-RL	&	0.14	&	0.32	&	0.46	&	0.50	&	1.00	&	0.93	&	1.19	\\
&	&	AL	&	0.35	&	0.19	&	0.54	&	0.47	&	1.00	&	0.53	&	1.18	\\
&	&	$L_1$-AL	&	0.28	&	0.37	&	0.64	&	0.52	&	1.00	&	0.86	&	1.18	\\
&	&	QL	&	0.35	&	0.16	&	0.41	&	0.43	&	1.00	&	0.30	&	1.17	\\
&	&	$L_1$-QL	&	0.16	&	0.28	&	0.44	&	0.46	&	1.00	&	0.41	&	1.19	\\
\cline{2-10}
&\multirow{8}{*}{30}		&	MCMEA	&	0.35	&	0.12	&	0.47	&	0.51	&	1.00	&	0.54	&	1.18	\\
&	&	$L_1$-MCMEA	&	0.28	&	0.26	&	0.54	&	0.55	&	1.00	&	0.99	&	1.18	\\
&	&	RL	&	0.20	&	0.26	&	0.46	&	0.50	&	1.00	&	0.60	&	1.19	\\
&	&	$L_1$-RL	&	0.19	&	0.32	&	0.51	&	0.53	&	1.00	&	0.98	&	1.19	\\
&	&	AL	&	0.39	&	0.19	&	0.58	&	0.50	&	1.00	&	0.58	&	1.18	\\
&	&	$L_1$-AL	&	0.30	&	0.38	&	0.68	&	0.55	&	0.99	&	0.99	&	1.18	\\
&	&	QL	&	0.30	&	0.15	&	0.45	&	0.46	&	0.99	&	0.97	&	1.13	\\
&	&	$L_1$-QL	&	0.16	&	0.32	&	0.48	&	0.49	&	0.99	&	0.99	&	1.19	\\
\cline{2-10}
&\multirow{8}{*}{50}		&	MCMEA	&	0.37	&	0.15	&	0.52	&	0.54	&	1.00	&	0.60	&	1.17	\\
&	&	$L_1$-MCMEA	&	0.29	&	0.29	&	0.58	&	0.58	&	1.00	&	0.99	&	1.18	\\
&	&	RL	&	0.21	&	0.27	&	0.48	&	0.53	&	1.00	&	0.61	&	1.17	\\
&	&	$L_1$-RL	&	0.21	&	0.35	&	0.56	&	0.56	&	1.00	&	0.99	&	1.19	\\
&	&	AL	&	0.42	&	0.20	&	0.62	&	0.53	&	1.00	&	0.63	&	1.16	\\
&	&	$L_1$-AL	&	0.32	&	0.40	&	0.72	&	0.58	&	1.00	&	0.99	&	1.18	\\
&	&	QL	&	0.32	&	0.17	&	0.49	&	0.49	&	1.00	&	0.97	&	1.09	\\
&	&	$L_1$-QL	&	0.20	&	0.32	&	0.52	&	0.52	&	0.98	&	1.00	&	1.18	\\
\hline
\end{tabular}

  \end{threeparttable}
\end{table}

\subsubsection*{B.2. Additional simulation settings and results}

\noindent\textbf{B.2.1. High-dimensional training sample with nonparametric independence screening} 

We designed and conducted simulation on an additional parameter setting, to assess the performance of the proposed methods in high-dimension situations. As described in Section~2.3, we used NIS to screen the covariates in the first step. 

We generated data as follows. the dimension of the covariates was indexed by $p$: 
$$\begin{aligned}
&\mathbf{X}_i \sim N_p(0,\mathbf{\Sigma}), \quad diag(\mathbf{\Sigma})=\mathbf{1}, \quad Corr(X_{ij},X_{ik})=0.5^{|j-k|}, i=1,...,n,\\
&D_i|\mathbf{X}_i \sim Bernoulli(p(\mathbf{X}_i)), \quad T_i = 2D_i-1, \quad logit(p(\mathbf{X}_i))=X_{i1}-X_{i2},\\
&Y_i = b_0(\mathbf{X}_i) + \frac{T_i}{2} \tau_0(\mathbf{X}_i)+\varepsilon_i, \quad \varepsilon_i \sim (1-p_o)N(0,1)+p_o Laplace(0,10), \\
&b_0(\mathbf{X}_i)= 0.5+4 X_{i1} +X_{i2}-3X_{i3},
\tau_0(\mathbf{X}_i)=2sin(2X_{i1})-X_{i2}+3tanh(0.5X_{i3}),
\end{aligned}$$
where $p_o$ is the proportion of outliers, $n=1000$, $p_o \in \{0,0.05\}$, and $p \in \{1000, 3000, 5000\}$

Figure B.1 shows that the NIS performed well in variable selection, especially when there were outliers. 
\begin{figure}[H]
\centering
\includegraphics[scale=0.6]{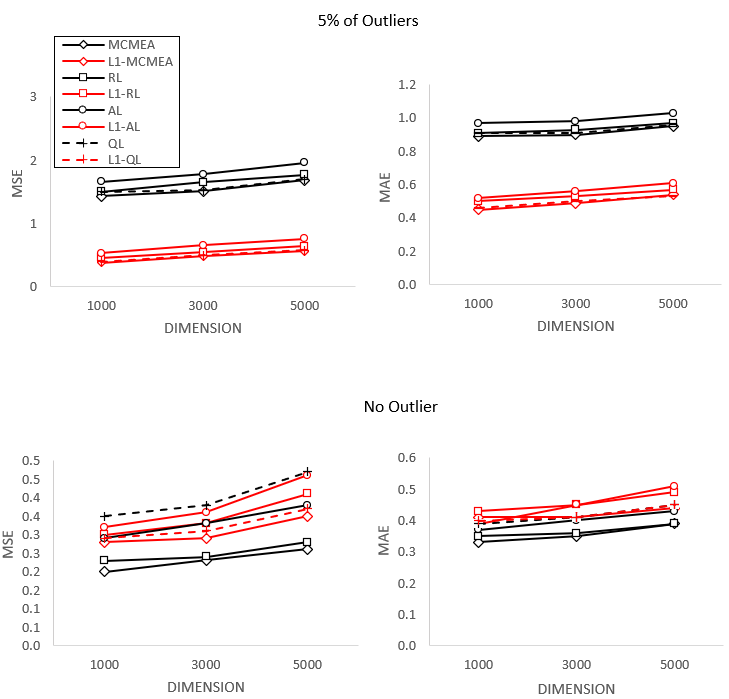}
\begin{minipage}[t]{0.8\textwidth} 
\caption*{Figure B.1. Influence of dimension. MSE and MAE tended to increase with dimension. But when there were outliers, the $L_1$-based methods (red line) performed markedly better than the $L_2$-based methods (black line).
}
\label{setup4}
\end{minipage}
\end{figure}

\begin{table}[H]
\begin{threeparttable}
\caption*{Table B.4: Simulation results of Setting 4. In the presence of outliers, the $L_1$-based methods performed markedly better than the $L_2$-based methods}
\begin{tabular}{ |l|l|l|c|c|c|c|c|c|c| } 
\hline
$\xi_o$ & Dimension	&Method		&	Bias.sq	&	Var	&	MSE	&	MAE	&	Sensitivity	&	Specificity & $Q(\hat{\eta})$	\\
\hline\hline
\multirow{24}{*}{0.05}&	\multirow{8}{*}{1000}& MCMEA	&	0.65	&	0.78	&	1.43	&	0.89	&	0.91	&	1.00	&	1.07	\\
&	&	$L_1$-MCMEA	&	0.07	&	0.32	&	0.38	&	0.45	&	0.98	&	1.00	&	1.18	\\
&	&	RL	&	0.61	&	0.88	&	1.50	&	0.91	&	0.89	&	1.00	&	1.06	\\
&	&	$L_1$-RL	&	0.06	&	0.40	&	0.46	&	0.50	&	0.98	&	1.00	&	1.18	\\
&	&	AL	&	0.88	&	0.78	&	1.66	&	0.97	&	0.86	&	1.00	&	1.03	\\
&	&	$L_1$-AL	&	0.12	&	0.41	&	0.53	&	0.52	&	0.98	&	1.00	&	1.17	\\
&	&	QL	&	0.78	&	0.71	&	1.50	&	0.91	&	0.94	&	1.00	&	1.07	\\
&	&	$L_1$-QL	&	0.09	&	0.32	&	0.40	&	0.46	&	0.98	&	1.00	&	1.18	\\
\cline{2-10}
&\multirow{8}{*}{3000}	&	MCMEA	&	0.56	&	0.94	&	1.51	&	0.90	&	0.88	&	1.00	&	1.05	\\
&	&	$L_1$-MCMEA	&	0.07	&	0.41	&	0.48	&	0.49	&	0.95	&	1.00	&	1.17	\\
&	&	RL	&	0.52	&	1.13	&	1.65	&	0.93	&	0.87	&	1.00	&	1.04	\\
&	&	$L_1$-RL	&	0.07	&	0.48	&	0.55	&	0.53	&	0.95	&	1.00	&	1.17	\\
&	&	AL	&	0.77	&	1.01	&	1.78	&	0.98	&	0.84	&	1.00	&	1.02	\\
&	&	$L_1$-AL	&	0.15	&	0.51	&	0.66	&	0.56	&	0.95	&	1.00	&	1.16	\\
&	&	QL	&	0.67	&	0.85	&	1.52	&	0.91	&	0.92	&	1.00	&	1.05	\\
&	&	$L_1$-QL	&	0.11	&	0.39	&	0.50	&	0.50	&	0.95	&	1.00	&	1.17	\\
\cline{2-10}
&\multirow{8}{*}{5000}	&	MCMEA	&	0.72	&	0.96	&	1.68	&	0.95	&	0.86	&	1.00	&	1.04	\\
&	&	$L_1$-MCMEA	&	0.13	&	0.43	&	0.56	&	0.54	&	0.93	&	1.00	&	1.16	\\
&	&	RL	&	0.67	&	1.10	&	1.77	&	0.97	&	0.85	&	1.00	&	1.04	\\
&	&	$L_1$-RL	&	0.12	&	0.52	&	0.64	&	0.57	&	0.93	&	1.00	&	1.16	\\
&	&	AL	&	0.98	&	0.98	&	1.96	&	1.03	&	0.82	&	1.00	&	1.01	\\
&	&	$L_1$-AL	&	0.24	&	0.52	&	0.76	&	0.61	&	0.93	&	1.00	&	1.15	\\
&	&	QL	&	0.86	&	0.84	&	1.70	&	0.96	&	0.91	&	1.00	&	1.05	\\
&	&	$L_1$-QL	&	0.17	&	0.42	&	0.58	&	0.53	&	0.93	&	1.00	&	1.16	\\
\hline
\multirow{24}{*}{0}&\multirow{8}{*}{1000}	&	MCMEA	&	0.10	&	0.10	&	0.20	&	0.33	&	1.00	&	1.00	&	1.21	\\
&	&	$L_1$-MCMEA	&	0.06	&	0.22	&	0.28	&	0.41	&	1.00	&	1.00	&	1.20	\\
&	&	RL	&	0.10	&	0.13	&	0.23	&	0.35	&	1.00	&	1.00	&	1.20	\\
&	&	$L_1$-RL	&	0.08	&	0.28	&	0.30	&	0.43	&	1.00	&	1.00	&	1.20	\\
&	&	AL	&	0.17	&	0.12	&	0.29	&	0.37	&	1.00	&	1.00	&	1.19	\\
&	&	$L_1$-AL	&	0.11	&	0.21	&	0.32	&	0.39	&	1.00	&	1.00	&	1.19	\\
&	&	QL	&	0.22	&	0.14	&	0.35	&	0.39	&	1.00	&	1.00	&	1.18	\\
&	&	$L_1$-QL	&	0.09	&	0.20	&	0.29	&	0.40	&	1.00	&	1.00	&	1.20	\\
\cline{2-10}
&\multirow{8}{*}{3000}	&	MCMEA	&	0.12	&	0.10	&	0.23	&	0.35	&	1.00	&	1.00	&	1.21	\\
&	&	$L_1$-MCMEA	&	0.06	&	0.23	&	0.29	&	0.41	&	1.00	&	1.00	&	1.20	\\
&	&	RL	&	0.11	&	0.13	&	0.24	&	0.36	&	1.00	&	1.00	&	1.20	\\
&	&	$L_1$-RL	&	0.06	&	0.28	&	0.33	&	0.45	&	1.00	&	1.00	&	1.20	\\
&	&	AL	&	0.20	&	0.13	&	0.33	&	0.40	&	1.00	&	1.00	&	1.19	\\
&	&	$L_1$-AL	&	0.10	&	0.26	&	0.36	&	0.45	&	1.00	&	1.00	&	1.19	\\
&	&	QL	&	0.22	&	0.15	&	0.38	&	0.41	&	1.00	&	1.00	&	1.18	\\
&	&	$L_1$-QL	&	0.08	&	0.23	&	0.31	&	0.41	&	1.00	&	1.00	&	1.20	\\
\cline{2-10}
&\multirow{8}{*}{5000}	& MCMEA	&	0.15	&	0.11	&	0.26	&	0.39	&	1.00	&	1.00	&	1.20	\\
&	&	$L_1$-MCMEA	&	0.11	&	0.24	&	0.35	&	0.28	&	1.00	&	1.00	&	1.20	\\
&	&	RL	&	0.14	&	0.13	&	0.28	&	0.39	&	1.00	&	1.00	&	1.20	\\
&	&	$L_1$-RL	&	0.11	&	0.30	&	0.41	&	0.49	&	1.00	&	1.00	&	1.20	\\
&	&	AL	&	0.25	&	0.13	&	0.38	&	0.43	&	1.00	&	1.00	&	1.19	\\
&	&	$L_1$-AL	&	0.18	&	0.28	&	0.46	&	0.51	&	1.00	&	1.00	&	1.19	\\
&	&	QL	&	0.30	&	0.17	&	0.47	&	0.45	&	1.00	&	1.00	&	1.18	\\
&	&	$L_1$-QL	&	0.13	&	0.24	&	0.37	&	0.45	&	1.00	&	1.00	&	1.19	\\

\hline
\end{tabular}

  \end{threeparttable}
\end{table}

\noindent\textbf{B.2.2. Effects of Smoothness Penalty}

Finally, we  investigated the effects of an added smoothness penalty. We considered a situation involving a univariate covariate $x$. We visualized the performance differences of the $L_1$ and $L_2$ methods, with and without the smoothness penalty. 

We generated the data of Setting 5 as follows: $X_i \sim Unif(0,1), \tau(X_i)=3sin(9(X_i-0.5)), p(X_i) = 1/(1+e^{-X_i}), Y_i=1+\frac{T_i}{2}\tau_0(X_i)+\varepsilon_i, \varepsilon_i\sim 0.9N(0,1)+0.1logNormal(0,4)$, the sample size $n=1000$, and the validation set with a size of $200$.  From Figures B.2 and B.3, it is clear that the $L_1$ methods outperform the $L_2$ methods, and the $L_1$ with smoothness penalty greatly  improved the performance of the estimation while reducing the variance. The 95\% Bootstrap C.I. coverage rate at selected point $x\in\{0.2,0.5,0.8\}$ were listed in Table B.5. The asymptotic variances of the $L_1$-MCM-EA methods without penalty at selected point $x\in\{0.2,0.5,0.8\}$ are 0.130, 0.110, and 0.137, corresponding 95\% asymptotic C.I. coverage rates are 0.957, 0.973, and 0.959. The asymptotic variances of the $L_1$-RL methods without penalty at selected points are 0.129, 0.110, and 0.130, corresponding 95\% asymptotic C.I. coverage rates are 0.953, 0.969, and 0.957.

\begin{figure}[H]
\centering
\includegraphics[scale=0.7]{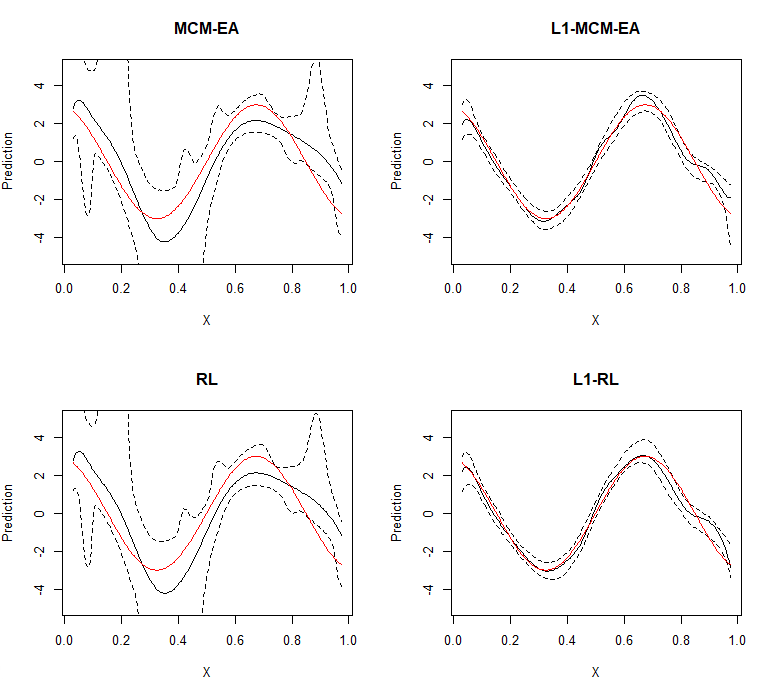}
\begin{minipage}[t]{0.8\textwidth} 
\caption*{Figure B.2. Panels on the left are $L_2$-based methods with smoothness penalty. Panels on the right are $L_1$-based methods, also with smoothness penalties. The black solid line is the estimate from one replication, the black dashed lines represent quantile 95\% bootstrap confidence interval from the same replication, and the red solid line represents the true treatment effect function.}
\label{setup5.1}
\end{minipage}
\end{figure}

\begin{figure}[H]
\centering
\includegraphics[scale=0.7]{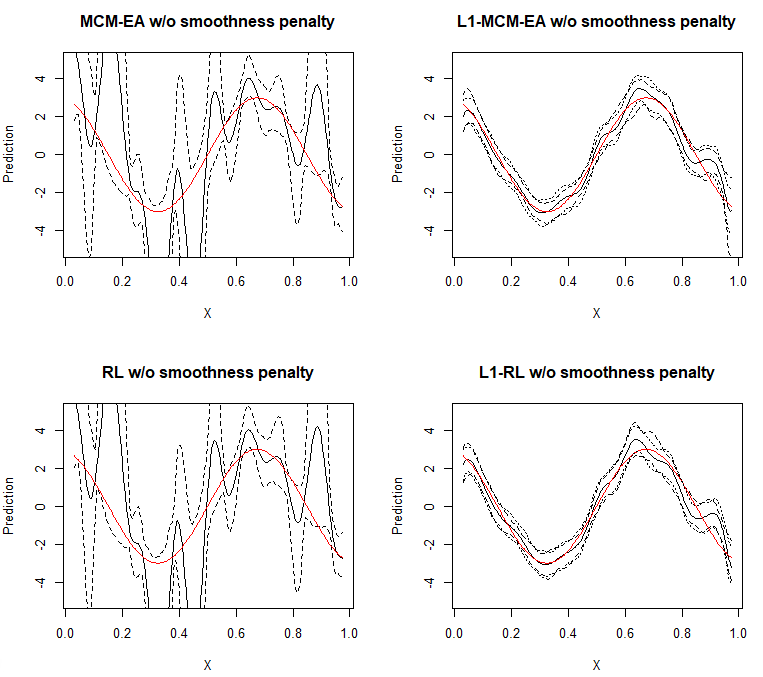}
\begin{minipage}[t]{0.8\textwidth} 
\caption*{Figure B.3. Panels on the left are $L_2$ loss based methods without smoothness penalties, panels on the right are $L_1$- based methods without smoothness penalties. The black solid line is the estimate from one replication, the black dashed lines represent quantile 95\% bootstrap C.I., the black dotted lines represent the 95\% asymptotic C.I. from the same replication, and the red solid line represents the true treatment effect function.}
\label{setup5.2}
\end{minipage}
\end{figure}

\begin{table}[H]
\caption*{Table B.5. Simulation results of Setting 5. The $L_1$-based methods generally produced coverage probabilities very close to the nominal level, even with the presence of outliers, whereas the $L_2$-based methods’ coverage sometimes deviated strongly from 0.95.}
\label{smooth}
\begin{center}
\begin{tabular}{ |l|l|l|c| }
\hline
Method & X & $\tau(X)$ & \shortstack{95\% Bootstrap C.I.\\ Coverage Rate} \\
\hline
MCMEA & 0.2 & -1.28 & 0.773 \\ 
& 0.5 & 0 &0.955 \\
& 0.8 & 1.28 &  0.791 \\
\hline
$L_1$-MCMEA & 0.2 & -1.28 &  0.954 \\ 
& 0.5 & 0 &  0.956 \\
& 0.8 & 1.28 &  0.945 \\
\hline
RL & 0.2 & -1.28 &  0.784 \\ 
& 0.5 & 0 &  0.962 \\
& 0.8 & 1.28 &  0.798 \\
\hline
$L_1$-RL & 0.2 & -1.28 &  0.950 \\ 
& 0.5 & 0 &  0.951 \\
& 0.8 & 1.28 &  0.951 \\
\hline
MCMEA w/o penalty & 0.2 & -1.28 &  0.927  \\ 
& 0.5 & 0 &  0.928 \\
& 0.8 & 1.28 &  0.931 \\
\hline
$L_1$-MCMEA w/o penalty & 0.2 & -1.28  & 0.955 \\ 
& 0.5 & 0 &  0.968 \\
& 0.8 & 1.28 &  0.963\\
\hline
RL w/o penalty & 0.2 & -1.28 &  0.929  \\ 
& 0.5 & 0 &  0.926 \\
& 0.8 & 1.28 &  0.931 \\
\hline
$L_1$-RL w/o penalty & 0.2 & -1.28 & 0.959 \\
& 0.5 & 0 & 0.969 \\
& 0.8 & 1.28 & 0.962 \\
\hline
\end{tabular}
\end{center}
\end{table}

\noindent\textbf{B.2.3. Comparison of Q-learning and A-learning when model is mis-specified}

We investigated the performance of Q-learning and proposed methods when model is mis-specified. We summarize the MSE and MAE of the Q-learning and proposed methods with combination of $L_1$ and $L_2$ loss when there is a small amount of outliers.

We generated the data for Setting 6 as follows: 
$$\begin{aligned}
&\mathbf{X}_i \sim N_p(0,\mathbf{\Sigma}),  diag(\mathbf{\Sigma})=\mathbf{1}, Corr(X_{ij},X_{ik})=0.5^{|j-k|}, i=1,...,n,\\
&D_i|\mathbf{X}_i \sim Bernoulli(p(\mathbf{X}_i)), T_i = 2D_i-1, logit(p(\mathbf{X}_i))=X_{i1}-X_{i2},\\
&Y_i = b_0(\mathbf{X}_i) + \frac{T_i}{2} \tau_0(\mathbf{X}_i)+\varepsilon_i,\varepsilon_i \sim (1-\xi_o)N(0,1)+\xi_o Laplace(0,10), \\
&b_0(\mathbf{X}_i)= 0.5+X_{i1} +X_{i2}^2-6X_{i3},
\tau_0(\mathbf{X}_i)=2sin(2X_{i1})-X_{i2}+3tanh(0.5X_{i3}),
\end{aligned}$$
where $n=1000$, $q=10$, $\xi_o=0.1$, and $Q(\eta^{opt})=2.18$. 
For Q-learning, the objective function is 
$$L_n(\boldsymbol\beta)=\frac{1}{n}\sum_{i=1}^n \rho\left(Y_i-\mathbf{X}_i^T\boldsymbol\gamma-\frac{T_i}{2}B(\mathbf{X}_i)^T\boldsymbol\beta\right)+\Lambda_n(\boldsymbol\beta),$$
where we used $L_1$ or $L_2$ loss functions for $\rho$. The results are summarised in Table B.6.

\begin{table}[H]
\caption*{Table B.6. Simulation results of Setting 6. In the presence of outliers, bias in the mis-specified $L_1$-QL was larger than that of the $L_1$-MCMEA, $L_1$-RL,and $L_1$-AL. The same was also true for MSE and MAE.}
\label{smooth2}
\begin{center}
\begin{tabular}{ |l|l|l|l|l|l|l|c| }
\hline
Method & Bias.sq  & Var & MSE & MAE & Sensitivity & Specificity & $Q(\hat{\eta})$ \\
\hline
MCMEA & 1.50 & 0.82 & 2.32 & 1.21 & 0.68 & 0.63 & 1.85\\ 
\hline
$L_1$-MCMEA & 0.18 & 0.64 &  0.82 & 0.62 & 0.99 & 0.74 & 2.14\\ 
\hline
RL & 1.57 & 0.73 &  2.30 & 1.21 & 0.68 & 0.65 & 1.86\\ 
\hline
$L_1$-RL & 0.19 & 0.63 &  0.83 & 0.61 & 0.99 & 0.73 & 2.15\\ 
\hline
AL & 1.43 & 0.89 &  2.32 & 1.20 & 0.69 & 0.65 & 1.87 \\ 
\hline
$L_1$-AL & 0.35 & 2.25  & 2.61 & 0.84 & 0.94 & 0.75 & 2.12\\ 
\hline
QL & 1.66 & 1.11 & 2.76 & 1.31 & 0.74 & 0.63 & 1.81 \\ 
\hline
$L_1$-QL & 3.57 & 1.01 & 4.58 & 1.15 & 1.00 & 0.52 & 2.08\\
\hline
\end{tabular}
\end{center}
\end{table}

\newpage
\subsection*{C. Real Data Application}
\subsubsection*{C.1. Existence of outliers}
The following figure shows that the outcome observations in both treatment groups are beyond normally distributed. 
\begin{figure}[!ht]
\centering
\includegraphics[scale=0.7]{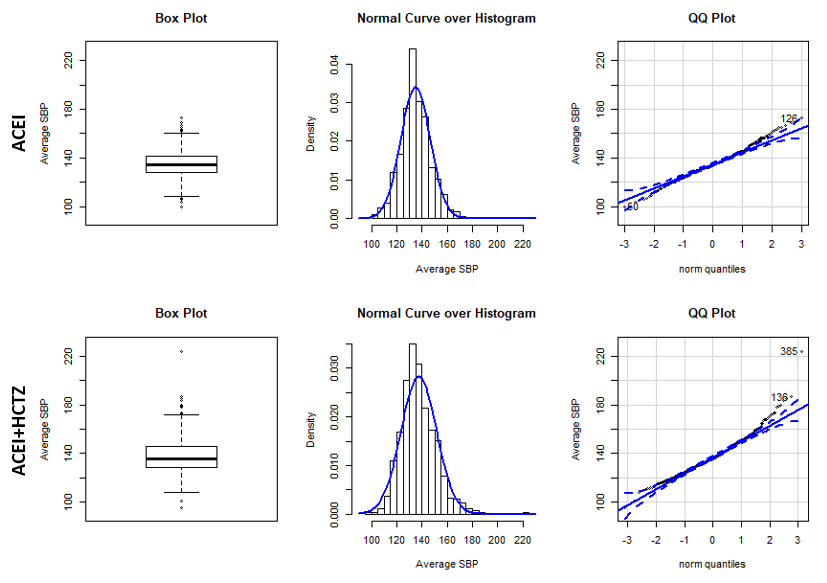}
\caption*{Figure C.1: Heavy-tailed Systolic Blood Pressure Distribution}
\label{heavytail}
\end{figure}

\subsubsection*{C.2: Nuisance quantity estimation}
The GBM is used to estimate mean outcome and propensity score. The estimation of two groups are as following figure.
\begin{figure}[H]
\centering
\includegraphics[scale=0.6]{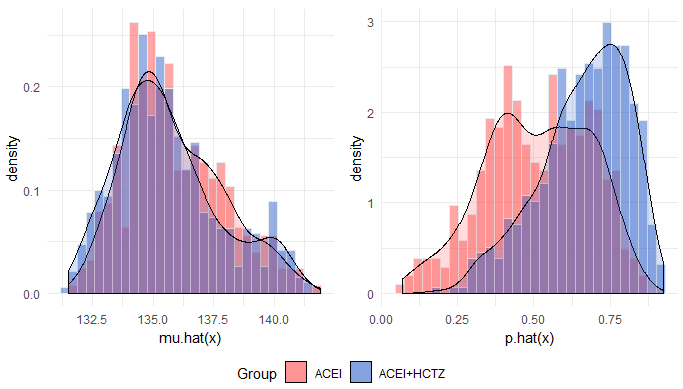}
\caption*{Figure C.2: Histograms of the mean outcomes and the estimated propensity score in the two treatment groups. The mean functions had similar shapes whereas the propensity distributions were clearly different.}
\label{propensity}
\end{figure}

In the application of proposed method, the importance levels from GBM are consistent with the result from regression. The importance levels from GBM and the linear and logistic regression results are summarized in the following two tables.
\begin{table}[H]
\caption*{Table C.1: Importance levels from the GBM analysis vs coefficients and p-values from regression analysis. }
\begin{center}
\begin{tabular}{ |l|c|c|c| } 
\hline
Variable &	Importance (scaled) &	Linear regression coefficient &	Linear regression p-value \\
\hline
Average PDC &	100.0000 &	-8.3352 &	<0.001*\\
BMI &	71.4317 &	0.1217 &	0.022*\\
Pulse &	57.9355 &	0.0829 &	0.052\\
Male &	51.0437 &	5.3345 &	<0.001*\\
Age &	40.9329 &	0.1421 &	<0.001*\\
Depression &	21.7349 &	-2.7787 &	0.007*\\
CAD &	9.6647 &	3.9908 &	0.181\\
Diabetes &	8.3385 &	-1.4464 &	<0.001*\\
Stroke &	5.7007 &	3.9820 &	0.242\\
Hyperlipidemia &	3.3518 &	-0.6630 &	0.581\\
Black &	2.9593 &	0.8341 &	0.351\\
CKD &	1.6875 &	-4.6852 &	0.101\\
COPD &	0.7970 &	-1.6243 &	0.272\\
CHF &	0.3214 &	1.4062 &	0.668\\
Atrial fibrillation &	0 &	0.2215 &	0.968\\
MI &	0 &	-5.3773 &	0.405\\
\hline
\end{tabular}
\end{center}
\end{table}

\begin{table}[H]
\caption*{Table C.2: Propensity models based on GBM and logistic regression}
\begin{center}
\begin{tabular}{ |l|c|c|c| } 
\hline
Variable &	Importance (scaled) &	Logistic regression coefficient &	Logistic regression p-value \\
\hline
BMI &	100.0000 &	0.0342 &	<0.001*\\
Pulse &	75.3721 &	-0.0162 &	0.028*\\
Age &	65.0922 &	0.0205 &	0.002*\\
Diabetes &	61.9452 &	-1.2126 &	<0.001*\\
Black &	27.7566 &	0.7032 &	<0.001*\\
Male &	8.5503 &	-0.2410 &	0.123\\
Hyperlipidemia &	5.2229 &	0.3572 &	0.091\\
Depression &	3.3688 &	0.2294 &	0.200\\
COPD &	2.6379 &	-0.2325 &	0.353\\
CAD &	2.0312 &	0.2867 &	0.577\\
Stroke &	1.9077 &	-0.9878 &	0.086\\
CKD &	1.1372 &	0.1413 &	0.772\\
CHF &	0.5027 &	0.1182 &	0.834\\
MI &	0 &	0.4516 &	0.678\\
Atrial fibrillation &	0 &	1.5788 &	0.176\\
\hline
\end{tabular}
\end{center}
\end{table}

\end{document}